\documentclass[10pt,journal,compsoc]{IEEEtran}

\usepackage{mathrsfs}
\usepackage{amsfonts}
\usepackage{epsfig,graphics,color}

\usepackage[shortlabels]{enumitem}
\usepackage{algorithm, algorithmic}
\usepackage{tikz}
\usetikzlibrary{shapes,arrows}
\usepackage{ifpdf}
\usepackage{cite}
\ifCLASSINFOpdf

\else \fi
\usepackage[cmex10]{amsmath}
\usepackage{array}
\usepackage{mdwmath}
\usepackage{mdwtab}
\usepackage{eqparbox}
\usepackage[tight,footnotesize]{subfigure}
\usepackage{amsmath}

\usepackage{amssymb}
\usepackage{amsmath}   % add1
\usepackage{booktabs}  % add2
\usepackage{threeparttable} %add3
\usepackage{multirow}       %add4

\ifCLASSOPTIONcompsoc
\def\IEEEQED{\IEEEQEDclosed}   % default to closed for compsoc
\else
\def\IEEEQED{\IEEEQEDopen} % otherwise default to open
\fi
\usepackage{amsthm}

\theoremstyle{plain}
\newtheorem{theorem}{\textbf{Theorem}}

\newtheorem{remark}{\textbf{Remark}}

\usepackage{ragged2e}

\renewcommand{\raggedright}{\leftskip=0pt \rightskip=0pt plus 0cm}

\begin{document}
\title{Optimal Status Updates for Minimizing Age of Correlated Information in IoT Networks with Energy Harvesting Sensors}

\author{
Chao Xu, {\em Member, IEEE}, Xinyan Zhang, Howard H. Yang, {\em Member, IEEE},  Xijun Wang, {\em  Member, IEEE},\\ Nikolaos Pappas, {\em Senior Member, IEEE}, Dusit Niyato, {\em Fellow, IEEE}, and Tony Q. S. Quek, {\em Fellow, IEEE}% <-this % stops a space
\IEEEcompsocitemizethanks{
\IEEEcompsocthanksitem C. Xu and X. Zhang are with the School of Information Engineering, Northwest A\&F University, Shaanxi, China, and also with the Key Laboratory of
Agricultural Internet of Things, Ministry of Agriculture and Rural Affairs, Northwest A\&F University, Yangling 712100, China (e-mail: \{cxu, zhangxinyan\}@nwafu.edu.cn).
\IEEEcompsocthanksitem H. H. Yang is with the Zhejiang University/University of Illinois at Urbana-Champaign Institute, Zhejiang University, Haining, China (email: haoyang@intl.zju.edu.cn).
\IEEEcompsocthanksitem X. Wang is with the School of Electronics and Information Technology, SYSU, Guangzhou, China (e-mail: \mbox{wangxijun@mail.sysu.edu.cn}).
\IEEEcompsocthanksitem N. Pappas is with the Department of Computer and Information Science, Link\"{o}ping University, Link\"{o}ping, Sweden (e-mail: nikolaos.pappas@liu.se).
\IEEEcompsocthanksitem D. Niyato is with the School of Computer Science and Engineering, Nanyang Technological University, Singapore (e-mail: \mbox{DNIYATO@ntu.edu.sg}).
\IEEEcompsocthanksitem T. Q. S. Quek is with the Singapore University of Technology and Design, Singapore 487372, and also with the Yonsei Frontier Lab, Yonsei University, South Korea (e-mail: tonyquek@sutd.edu.sg).
}% <-this % stops an unwanted space
\thanks{This paper is supported by the National Natural Science Foundation of China (62271413, 62201504, 62271513), the Research Fund under the Shaanxi Province Innovation Capability Support Program (2023KJXX-010), the Guangdong Basic and Applied Basic Research Foundation (2021A1515012631), the Zhejian
g Provincial Natural Science Foundation of China (LGJ22F010001), the Zhejiang-Singapore Innovation and AI Joint Research Lab, the National Research Foundation, Singapore and Infocomm Media Development Authority under its Future Communications Research \& Development Programme, the DSO National Laboratories under the AI Singapore Programme (AISG2-RP-2020-019), the MOE Tier 1 (RG87/22), the MOE ARF Tier 2 (T2EP20120-0006), the Swedish Research Council (VR), ELLIIT, and the European Union (ETHER, 101096526). This paper was presented in part in the Proc. of WCSP \cite{AoCI_Xu_EH_WCSP}. (Corresponding Author: Howard H. Yang, Xijun Wang, and T. Q. S. Quek)}
}

\IEEEtitleabstractindextext{%
\begin{abstract}
\raggedright {Many real-time applications of the Internet of Things (IoT) need to deal with correlated information generated by multiple sensors. The design of efficient status update strategies that minimize the Age of Correlated Information (AoCI) is a key factor. In this paper, we consider an IoT network consisting of sensors equipped with the energy harvesting (EH) capability.
We optimize the average AoCI at the data fusion center (DFC) by appropriately managing the energy harvested by sensors, whose true battery states are unobservable during the decision-making process. Particularly, we first formulate the dynamic status update procedure as a partially observable Markov decision process (POMDP), where the environmental dynamics are unknown to the DFC.
In order to address the challenges arising from the causality of energy usage, unknown environmental dynamics, unobservability of sensors' true battery states, and large-scale discrete action space, we devise a deep reinforcement learning (DRL)-based dynamic status update algorithm. The algorithm leverages the advantages of the soft actor-critic and long short-term memory techniques. Meanwhile, it incorporates our proposed action decomposition and mapping mechanism.
Extensive simulations are conducted to validate the effectiveness of our proposed algorithm by comparing it with available DRL algorithms for POMDPs.}
\end{abstract}

}

\maketitle

\IEEEdisplaynontitleabstractindextext

\IEEEpeerreviewmaketitle

\section{Introduction}\label{sec:introduction}
The rapid advancement of the Internet of Things (IoT) has brought a great influence to many areas and found numerous valuable real-time applications, ranging from high-speed assembly and packaging in factory automation, autonomous driving in intelligent transport systems, to environmental monitoring and process automation in precision agriculture \cite{Survey_IoT_Applications_2015,Real_Time_App_2017,IoT_Precise_Agriculture}.
For real-time applications, the Quality of Service (QoS) is strongly dependent on the timely delivery of information, since information staleness can incur severe degradation in the accuracy of the decision-making process.
Recently, Age of Information (AoI) has been proposed as an effective metric to quantify the timeliness of information deliveries \cite{AoI_Org_2012}, which has promoted new designs of status update strategies to improve the information freshness in IoT networks\cite{AoI_Survey_2017,Aoi_Survey_2021}.

A crucial challenge in providing fresh status updates for IoT networks is the limited energy supply of sensors, as they are usually powered by batteries with limited capacity rather than fixed power supplies \cite{Energy_Harvesting_AoI_Mag}.
Due to its cost-efficient implementation, the energy harvesting (EH) technique is emerging as a promising alternative to power IoT sensors. For sensors with EH capabilities, the status update policy shall be correspondingly adjusted to account for the harvested energy \cite{Aoi_Survey_2021,Energy_Harvesting_AoI_Mag}.
In previous researches on status update strategy design for EH sensors, it was commonly assumed that the information of sensors' battery states is fully available to the controller upon each time of decision making. Nevertheless, such an assumption is valid under the condition that the battery states of sensors can be readily known by the controller, which inevitably incurs excessive energy consumption and channel occupancy.

Meanwhile, for many IoT applications, status updates from different sensors are correlated and need to be integrated at the destination to produce useful decisions. Typical applications include the real-time temperature or air quality monitoring of a certain area (e.g., a city) using a set of sensors \cite{Corr_Tem_2021}, multimedia surveillance by deploying several cameras to perceive an environment from multiple viewpoints \cite{Carema_Net_2007}, and healthcare applications facilitated through the concurrent sampling of diverse processes across various scale spaces \cite{Correlation_Heallthcare_2010}.
In view of this, the concept of the Age of Correlated Information (AoCI) was introduced \cite{AoCI_WCN} to characterize the information freshness of correlated updates.
At the destination, AoCI increases linearly with time and decreases only when the \textit{newly integrated} information is generated. In order to minimize the AoCI, correlated sensors shall be jointly scheduled, since a piece of integrated information can be obtained only when the correlated update packets are ``fully" received.

Given the pervasiveness of the information correlation in practice \cite{Corr_Tem_2021,Carema_Net_2007,Correlation_Heallthcare_2010}, we devise a status update policy to optimize the AoCI in an IoT network consisting of a data fusion center (DFC) and a set of EH sensors that are monitoring multiple correlated sensing points (CSPs).
% , in which the AoCI is adopted to characterize the freshness of the integrated information at the data fusion center (DFC).
The sensing points are correlated in the sense that their status updates with the same timestamp need to be aligned at the DFC to produce a desired result, which can be further utilized by other real-time IoT applications.
% To be specific, we consider an IoT network consisting of one DFC, which integrates the status updates of CSPs, and a number of sensors monitoring the dynamics of these CSPs and refreshing the new status to the DFC.
Due to the constraints of the communication resources and energy causality, in each time slot only a subset of the sensors can be activated to sense the CSPs and upload the update packets to the DFC over error-prone links.
Besides, to avoid additional signaling overhead, we consider that the true battery states of sensors cannot be synchronously provided to the DFC.

To optimize the average AoCI at the DFC, it is essential to adequately schedule EH sensors to send timely status updates while fulfilling the EH and battery capacity constraints.{\footnote{Maximizing the energy efficiency of update transmissions can be another objective, which is beyond the scope of this study.} }
This problem is non-trivial due to ($a$) the causality of energy usage, ($b$) the unknown environmental dynamics, and ($c$) the unobservability of sensors' true battery states.
In this work, we cast the status update procedure as a partially observable Markov decision process (POMDP) \cite{POMDOP_Survey_2013}.
By jointly leveraging the soft actor-critic (SAC) \cite{SAC_Org_2018} and long short-term memory (LSTM) \cite{LSTM_1997} techniques, we develop a deep reinforcement learning (DRL)-based dynamic status update algorithm, coined as the Recurrent SAC Status update (RSS) algorithm.
To circumvent the challenge arising from the large-scale discrete action space, an action decomposition and mapping (ADM) mechanism has also been developed and embedded, by exploiting the structure of our studied problem.
Extensive simulations are conducted to verify the efficacy of our proposed algorithm over the baseline DRL algorithms. To the best of our knowledge, this is the first work that builds up RNN-enhanced DRL algorithms to optimize the AoCI for IoT networks with EH sensors.

Although the conference version \cite{AoCI_Xu_EH_WCSP} also studied an AoCI minimization problem, it had three main limitations:
1) the battery states of sensors were assumed to be observable to the DFC, and the considered problem was formulated as a Markov decision process (MDP);
2) the scale of the studied problem was relatively small and hence, only a deep Q-network (DQN)-based status update algorithm was developed to solve it;
and 3) simulations were not sufficient to evaluate the proposed algorithm's effectiveness where only random and greedy policies were used as the baseline.
This work overcomes the above-mentioned limitations and the main contributions can be summarized as follows.
% While our conference version \cite{AoCI_Xu_EH_WCSP} also aimed at minimizing the achieved average AoCI at the DFC, it had three main limitations: 1) The true battery states of all sensors were assumed to be observable by the DFC and the considered problem is formulated as a Markov decision process (MDP). 2) The concerned problem scale was relatively small, where only a relatively small-scale discrete action space (with no more than 200 valid actions) was incorporated in the formulated MDP. As such, a deep Q-network (DQN)-based status update algorithm was developed to address the formulated problem. 3) Simulations were not sufficient enough to evaluate the effectiveness of the proposed algorithm, since only the performance of the random and greedy policies were taken as the baseline.
% This work overcomes the aforementioned limitations, and the main contributions can be summarized as follows.
\begin{itemize} \setcounter{enumi}{0}
\item We formulate the considered dynamic status update procedure of EH sensors as a POMDP, in which the true battery states of sensors cannot be observed by the DFC at the time of decision making. To improve the stability of the solution's performance, we devise an optimal stochastic stationary policy, which is more beneficial for partially observed problems than the deterministic policy \cite{POMDP_SP_94,POMDP_SP_15}.
\item We exploit the advantages of the SAC and LSTM techniques and establish a DRL algorithm to solve the formulated POMDP problem, in which an ADM mechanism has been proposed and embedded. The proposed DRL algorithm is compatible with not only the large-scale state space, as available DRL algorithms (i.e., deep recurrent Q-network (DRQN) \cite{DRQN} and deep recurrent deterministic policy gradient (DRDPG) \cite{DRDPG_2015}), but also the large-scale discrete action space.
\item The effectiveness of our proposed RSS algorithm is verified by comparing it with several baseline DRL algorithms, i.e., DRQN, DRDPG with ADM (DRDPG-WA), DRDPG without ADM (DRDPG-WOA), and RSS without ADM (RSS-WOA).
    The extensive simulation results show that our algorithm has better convergence, scalability, and stability properties than the baseline algorithms in terms of the achieved average AoCI. More importantly, the convergence and stability of the RSS algorithm can be guaranteed even in the \textit{large-scale scenario with more than $800,000$ valid actions}.
\end{itemize}

The rest of the paper is organized as follows. In Section 2, we summarize the related work. In Section 3, the description of the system model and concerned problem are presented. In Section 4, we cast the dynamic status update procedure as a POMDP and then, develop the DRL algorithm to solve it. Extensive simulation results are presented to show the effectiveness of our proposed scheme in Section 5, and conclusions are drawn in Section 6.

\section{Related Work}

The notion of AoI has spurred a large body of research on dynamic status update strategy design for various IoT networks, e.g., AoI optimal status update for the sensor with and without EH, and timely status update for correlated information, which are briefly reviewed as follows.

\subsection{AoI optimal status update for the sensor without EH}
Recently, the AoI has been proposed as a new metric to assess the information freshness by measuring the time elapsed since the latest received packet was generated from the source \cite{AoI_Org_2012}. Armed with this metric, a few efforts began to investigate efficient status update strategies to improve information freshness in IoT networks \cite{AoI_Throughput_2019,BZhou_Samp_Up_2019,AoI_Pef_SCI,Our_IF_2019,Our_Caching_AoI_2021}. In \cite{AoI_Throughput_2019}, the authors developed status update policies to minimize the expected weighted sum AoI of the network subject to the throughput requirements of sensors.
Under an average cost constraint for sensors, the sampling and transmission processes were jointly optimized in \cite{BZhou_Samp_Up_2019} so as to minimize the average AoI at the destination.
In \cite{AoI_Pef_SCI}, an age-optimal update policy was developed for the IoT network with multiple transceiver pairs, where the channel states were assumed to be perfectly known by the scheduler.
An edge computing-enabled IoT network was considered in \cite{Our_IF_2019}, in which sensors' update rates were optimized to minimize the maximum average peak AoI. Authors in \cite{Our_Caching_AoI_2021} considered a caching-enabled IoT network, and devised a dueling deep R-network-based status update algorithm to balance the AoI experienced by users and energy consumption of sensors.

\subsection{AoI optimal status update for the EH sensor}
To make timely status updates in EH-enabled IoT networks, the fundamental question is how to efficiently manage the energy harvested by the sensor, which is nontrivial because of the energy causality constraint \cite{Aoi_Survey_2021,Energy_Harvesting_AoI_Mag}.
It has recently attracted increasing research interests \cite{EH_Model_TIT,Pappas_EH_AoI_1,AoI_Dis_IoT_2020,Pappas_EH_AoI_3,AoI_EH_Cach_Pappas_2021,EH_AoI_TMC_2021}.
Considering the IoT system with an EH sensor, authors in \cite{EH_Model_TIT} studied the optimality of threshold-based online status update policies for two different EH models. In \cite{Pappas_EH_AoI_1}, authors studied the optimal transmission policy for an EH status update system monitoring a stochastic process, which can be in a normal state or an alarm state.
For the IoT system consisting of a remote monitor and an EH-powered sensor, authors in \cite{AoI_Dis_IoT_2020} addressed the trade-off between the update timeliness and transmission distortion.
The wireless-powered IoT system with a transceiver pair was studied in \cite{Pappas_EH_AoI_3}, in which a status update policy was proposed to minimize the average AoI at the receiver.
For the caching-enabled IoT network with EH sensors, an AoI-oriented status update strategy was investigated in \cite{AoI_EH_Cach_Pappas_2021}.
In \cite{EH_AoI_TMC_2021}, an online algorithm was proposed to minimize the weighted sum of average peak AoI for an IoT network, where error-free channels were allocated in a round-robin fashion.

\subsection{Timely status update for correlated information}
While efficient status update policies have been proposed in the literature \cite{AoI_Pef_SCI,AoI_Throughput_2019,BZhou_Samp_Up_2019,Our_IF_2019,Our_Caching_AoI_2021,EH_Model_TIT,Pappas_EH_AoI_1,AoI_Dis_IoT_2020,Pappas_EH_AoI_3,AoI_EH_Cach_Pappas_2021,EH_AoI_TMC_2021}, it was commonly assumed that the update packets generated by different sensors were independent.
However, for many IoT applications, the update packets generated by sensors are correlated and contribute to the same decision-making process \cite{Corr_Tem_2021,Carema_Net_2007,Correlation_Heallthcare_2010}. Motivated by this, work \cite{AoCI_WCN} considered the system with a number of camera nodes (CNs) and studied the AoCI-oriented optimal association and transmission policy, but assuming that the update packets to be transmitted were pre-buffered in the CNs.
Considering the more general case with a generate-at-will model, RL/DRL-based dynamic status update schemes were proposed to improve the information freshness for correlated sensors \cite{APP_Level_Scheduling,AoCI_2_kinds_Sensor,Timely_Monitoring_2020}. Authors in \cite{APP_Level_Scheduling} extended \cite{AoCI_WCN} by dividing the time into frames, each consisting of a few time slots, and assumed that all sensors generated update packets at the beginning of each frame. Then, a DQN-based scheduling algorithm was developed to optimize the AoCI.
In \cite{AoCI_2_kinds_Sensor}, an age optimal scheduling scheme was proposed for the system with two types of correlated sensors, whose update arrivals were generated with the random and generated-at-will models, respectively.
It was assumed in \cite{AoCI_2_kinds_Sensor} that the destination could reconstruct the status information if the number of successfully delivered updates (regardless of their sources) in one time slot exceeded a certain threshold.
The work in \cite{Timely_Monitoring_2020} considered an IoT network with a set of sensing points, which could be simultaneously observed by sensors with different nonzero probabilities. Both the relative value iteration and heuristic scheduling schemes were proposed to optimize the information freshness at the destination in different cases. In \cite{APP_Level_Scheduling,AoCI_2_kinds_Sensor,Timely_Monitoring_2020}, no EH-powered sensors were considered.

\subsection{Summary of the comparison with related work}

%\begin{table}[!t]
%\centering
%\caption{A summary of the comparison between our study with related work.}
%\label{Tab:Comp_Related}
%\begin{tabular}{|c|c|c|c|}
%\hline
%\multicolumn{1}{|c|}{Work} & EH sensor & Correlated information & True battery state     \\ \hline
%\cite{AoI_Throughput_2019,BZhou_Samp_Up_2019,AoI_Pef_SCI,Our_IF_2019,Our_Caching_AoI_2021}                                &          & 16.48         & 2h 32m 37s           \\ \hline
%5                                & 0.44         & 16.58         & 2h 32m 51s          \\ \hline
%7                                & 0.44         & 16.76         & 2h 32m 58s           \\ \hline
%$K$                       & Parameters   & FLOPs          & Storage space (kB) \\ \hline
%3                                & 137,347      & 138,240        & 540                \\ \hline
%5                                & 140,677      & 141,568        & 554                \\ \hline
%7                                & 144,007      & 144,896        & 568                \\ \hline
%\end{tabular}
%\end{table}

%\textcolor{blue}{\textit{A summary of the comparison between our study with related work is shown in Table ****.}}
Unlike the above-mentioned researches \cite{AoI_Pef_SCI,AoI_Throughput_2019,BZhou_Samp_Up_2019,Our_IF_2019,Our_Caching_AoI_2021,EH_Model_TIT,Pappas_EH_AoI_1,AoI_Dis_IoT_2020,Pappas_EH_AoI_3,AoI_EH_Cach_Pappas_2021,EH_AoI_TMC_2021,APP_Level_Scheduling,AoCI_2_kinds_Sensor,Timely_Monitoring_2020}, we concentrate on devising a dynamic status update policy to minimize the AoCI for IoT networks with multiple CSPs, each of which can be observed by a set of EH sensors. Wherein two new challenges arise. On the one hand, in contrast to \cite{EH_Model_TIT,Pappas_EH_AoI_1,AoI_Dis_IoT_2020,Pappas_EH_AoI_3,AoI_EH_Cach_Pappas_2021}, we consider that the true battery states of sensors cannot be synchronously provided to the DFC at the decision-making time, as specified in Section 3.2, which avoids incurring extra signaling overhead (e.g., energy consumption and channel occupancy).
On the other hand, in contrast to previous studies on AoI optimization problems \cite{AoI_Pef_SCI,AoI_Throughput_2019,BZhou_Samp_Up_2019,Our_IF_2019,Our_Caching_AoI_2021,EH_Model_TIT,Pappas_EH_AoI_1,AoI_Dis_IoT_2020,Pappas_EH_AoI_3,AoI_EH_Cach_Pappas_2021,EH_AoI_TMC_2021}, we consider a scenario where multiple ``qualified'' (see Section 3.3) EH sensors need to be appropriately scheduled in each time slot, posing a significant challenge for solving the sequential decision-making problem with a large number of valid actions. Furthermore, the effects of EH sensors' status updates on the reward shall be jointly learned and cannot be readily decoupled. These challenges make the existing RL/DRL-based or the Whittle index-based scheduling approaches \cite{AoI_Pef_SCI,AoI_Throughput_2019,BZhou_Samp_Up_2019,Our_IF_2019,Our_Caching_AoI_2021,EH_Model_TIT,Pappas_EH_AoI_1,AoI_Dis_IoT_2020,Pappas_EH_AoI_3,AoI_EH_Cach_Pappas_2021,EH_AoI_TMC_2021,APP_Level_Scheduling,AoCI_2_kinds_Sensor,Timely_Monitoring_2020} inapplicable to our problem.
To this end, we formulate the concerned dynamic status update problem as a POMDP and develop a recurrent neural network (RNN)-enhanced DRL algorithm by utilizing the SAC, LSTM, and our proposed ADM techniques.

% \textcolor{red}{\textbf{Put the novelty here or at the end of Section 4.2.3 ?}}
The novelty of our proposed RSS algorithm lies in two aspects. First, to the best of our knowledge, this is the pioneering work that develops an RNN-enhanced DRL algorithm to solve POMDPs by simultaneously taking the advantages of SAC and LSTM techniques.
Particularly, we have defined the soft state- and action-value functions (i.e., (\ref{Eq:Act_Value_Fun}) and (\ref{Eq:State_Value_Fun})) regarding our formulated POMDP and theoretically established their recursive relationships (i.e., Theorem 1), which are the basis for deriving the expressions of the soft Bellman residual (SBR) and expected KL-divergence (EKLD) and training the constructed artificial neural networks (ANNs).
Besides, to improve the convergence and stability of the proposed RSS algorithm, the ANNs have been elaborately designed by fully considering features of our formulated POMDP, rather than simply following existing studies \cite{DRDPG_2015,POMDP_2017,MADRL_TMC_2021}.
Second, we have devised an ADM mechanism based on the structure of our concerned problem and embedded it in our proposed DRL algorithm to circumvent the challenge arising from the large-scale discrete action space. We note that the AMD mechanism is compatible with not only our proposed RSS algorithm but also other policy gradient DRL algorithms, e.g., the DRDPG. And, as shown in the simulation results in Section 5, with the increase of the action space size, the improvement in the convergence and stability brought by the ADM mechanism is more pronounced.

For ease of reference, we list the main notations used in this paper in Table \ref{notations_lable}.

\begin{table}[!t]
    \centering
    \renewcommand{\arraystretch}{1.2}
	\caption{Main notations used in this paper}
    \label{notations_lable}
	\begin{tabular}{|c|c|} \hline
	\textbf{Notation}                                 & \textbf{Description}                       \\ \hline
	$n, N, \mathcal{N}$                               & Index, number, set of EH sensors      \\ \hline
	$k, K, \mathcal{K}$                               & Index, number, set of CSPs               \\ \hline
	$\mathcal{N}_k$                                   & Set of EH sensors associated with CSP $k$               \\ \hline
%	$M$                                               & Number of orthogonal channels                \\ \hline
	$t$                                               & Index of time slot  \\ \hline
	$e_{n}(t)$                         & Battery state of EH sensor $n$ \\ \hline
    $\hat{e}_{n}(t)$                   & Residual energy informed by EH sensor $n$ \\ \hline
%	$\kappa_{n}(t)$                                   & Amount of energy harvested by EH sensor $n$ \\ \hline
	$D_n$                                             & Importance of data updated by EH sensor $n$                 \\ \hline
	$D_k^0$                                      & Importance threshold of CSP $k$                 \\ \hline
	% $\Upsilon_{k}(t), \Upsilon(t)$                    & .....               \\ \hline
	$\Delta(t)$                                           & AoCI at DFC               \\ \hline
	% $G_n(t), T_n(t), X_n(t)$                                            & .....               \\ \hline
	$\mathcal S(t),\mathcal O(t),U(t)$   & State, observation, reward            \\ \hline
	$\breve{\mathbf A}(t), \tilde{\mathbf A}(t), \mathbf{A}(t)$      & Primitive, proto, valid action                   \\ \hline
%	$\mathcal Z(t), \hat{\mathcal Z}(t)$              & Entire, approximated history        \\ \hline
	$\pi(\cdot), \pi^{*}(\cdot)$                                           & Policy, optimal policy              \\ \hline
	$V_{\pi}(\cdot), Q_{\pi}(\cdot, \cdot)$                 & State-, action-value function               \\ \hline
	$Q_{\theta_{j}}(\cdot, \cdot)$        & Critic network parameterized by $\theta_{j}$ \\ \hline
	$ Q_{\bar{\theta}_{j}}(\cdot, \cdot) $        & Target critic network parameterized by $\bar{\theta}_{j}$ \\ \hline
	$\pi_{\phi}(\cdot)$                                           & Actor network parameterized by $\phi$               \\ \hline
	\end{tabular}
\end{table}

\renewcommand\arraystretch{1.0}

\section{System Model and Problem Formulation} \label{Sec:Section 2}

\subsection{Network Model}
\begin{figure} [!t]
\centering \includegraphics[width=3.0in]{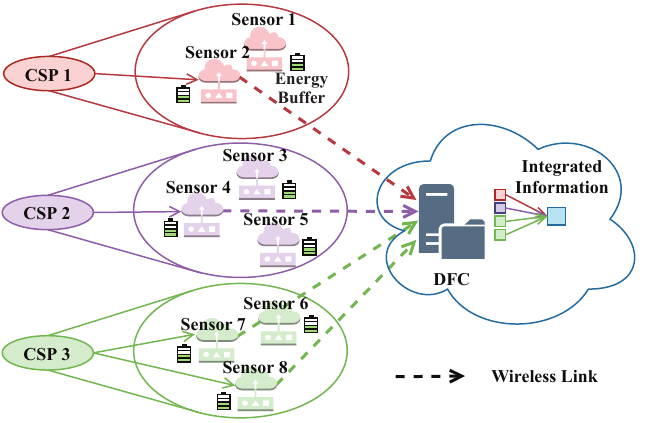}
\centering \caption{An illustration of the IoT network with EH sensors. The dashed lines denote orthogonal wireless channels.} \label{Fig:System_Model}
\end{figure}

As illustrated in Fig. \ref{Fig:System_Model}, we consider an IoT network consisting of $N$ EH sensors that are monitoring $K$ CSPs and sending the status data to a DFC.
The set of sensors and CSPs are denoted by $\mathcal N = \{1, 2, \ldots, N\}$ and $\mathcal K = \{1, 2, \ldots, K\}$, respectively.
Here, each sensor monitors only one CSP, and a status update packet with a timestamp is generated if the sensor is activated\footnote{Our established methodology can be extended to handle the more general case with some sensors being able to observe multiple CSPs, by introducing the concept of virtual sensors and appropriately modifying the proposed ADM mechanism.}.
The update packet will be transmitted to the DFC for further processing.
Let $\mathcal N_k$ denote the set of sensors monitoring CSP $k$, i.e., $\bigcup _{k=1}^K \mathcal N_k = \mathcal N, \mathcal N_k \bigcap \mathcal N_l = \varnothing, \forall k,l \in \mathcal K, k \neq l$.
We consider a discrete-time system, where time is divided into slots of unit length. At the beginning of each time slot, the DFC selects a subset of the sensors to sense the CSPs and then transmits their generated update packets.
This is known as the \textit{generate-at-will} model. For each sensor, it is assumed that the CSP sensing and subsequent update transmission take up one time slot.

At the end of each time slot, the successfully delivered update packets, if any, are processed by the DFC to produce the integrated information of interest.\footnote{The data processing time at the DFC is ignored here to ensure that the decision epochs are of uniform duration. Solving the problem with the non-uniform decision epoch is left for future work.}
Since wireless channels are unreliable, transmission failures may occur. We adopt a common assumption (see \cite{AoI_Throughput_2019,AoI_Pef_SCI,APP_Level_Scheduling,Timely_Monitoring_2020,Trans_Succ_Pro_TMC_2021,Our_Caching_AoI_2021} for instance) that the transmission failures of a generic sensor $n$ are independently and identically distributed over the time slots with a failure probability $p_n$, which is however unknown to the DFC and the sensor. Due to the limited communication resources, in each set $\mathcal N_k$ at most $M$ ($M \leq \left| {\mathcal N_k} \right|$) sensors are allowed to transmit update packets simultaneously without collisions (e.g., over orthogonal channels).
Let $\mathbf A(t) =  \left( \mathbf A_1(t),  \mathbf A_2(t), \ldots,  \mathbf A_K(t)\right)$ denote the scheduling decision made by the DFC in time slot $t$, with $\mathbf A_k = (A_n(t))_{n \in \mathcal N_k}$ and $A_{n}(t) \in \{0, 1\}, \forall n \in \mathcal N_k, \forall k \in \mathcal K$; $A_{n}(t)=1$ indicates that sensor $n$ is scheduled by the DFC to observe CSP $k$, and $A_{n}(t)=0$ otherwise. To this end, we establish the following relationship
\begin{align}  \label{equ:Sensing_Action_All}
\sum\nolimits_{n \in \mathcal N_k}  A_n(t) \leq  M, \forall k \in \mathcal K.
\end{align}

\subsection{Energy Model}
Each sensor $n$ is capable of harvesting energy from the ambient environment and storing the incoming energy in its battery, the capacity of which is $E_{n}$ units.
Following \cite{AoI_EH_Cach_Pappas_2021,Bernoulli_Unity_TMC_2018,Bernoulli_Unity_TMC_2018_2}, we assume that the energy required to complete one status update, including CSP sensing and data transmission, is normalized to one unit, i.e., $e_0 =1$. And, for a generic sensor $n$, the energy arrivals follow an independent Bernoulli process with parameter $\rho _{n}$.
It is worth noting that our model can also be extended to more general cases, where the sensors' power consumption is non-uniform or where the energy arrival process is modeled using other memoryless processes, such as Poisson, Markov, and correlated Bernoulli processes, as in \cite{EH_Model_TIT,Correlated_Bern_Tcom_2018}.

The energy harvested in time slot $t$ can be used only in future time slots, and a scheduled sensor can be activated only if there is enough energy in its battery. For sensor $n$, let $e_{n}(t) \in \mathbb \{0, 1, \ldots, E_{n}\}$ and $\hat A_{n}(t) \in \{0, 1\}$ respectively denote its battery state (i.e., the amount of available energy) at the beginning of and its operation state in time slot $t$, i.e., $\hat A_{n}(t) = 1$ if it is activated, and $\hat A_{n}(t) = 0$ otherwise. Then, the energy causality constraint is given by
\begin{align}  \label{Eq:Energy_Causality}
\hat e_{n}(t) = e_{n}(t) - \hat A_{n}(t)e_0 \geq 0
\end{align}
where $e_0=1$ denotes the energy consumption for completing one status update and $\hat{e}_{n}(t)$ the amount of residual energy at sensor $n$ after executing action $\hat A_{n}(t)$ without considering the energy arrival in this time slot.
%Note that assuming perfect knowledge of sensors' energy states at the DFC is not practical, because it is extremely costly, if not impossible, for the sensors to concurrently inform the DFC of their battery states at the beginning of each time slot.
Thereby, the relationship between $\hat A_{n}(t)$ and $A_{n}(t)$ can be expressed as
\begin{align}  \label{Eq:Activation_Condition}
\hat A_{n}(t) =
\begin{cases}
1, & \textrm{if} \ A_{n}(t) = 1  \textrm{ and} ~ \hat e_{n}(t) \geq 0\\
0, &  \textrm{otherwise}.
\end{cases}
\end{align}
%In other words, when a sensor is requested to update the status of a CSP, it only responds to the DFC if it has enough energy.
Moreover, the sensor's battery state evolves as
\begin{align}  \label{Eq:Battery_Evolve}
e_{n}(t) = \min\{\hat e_{n}(t-1) + \kappa_{n}(t-1), E_{n}\}
\end{align}
which is initialized as $e_{n}(1) = E_{n}, \forall n \in \mathcal N$. In (\ref{Eq:Battery_Evolve}), $\kappa_{n}(t-1) \in \{0, 1\}$ denotes the amount of energy harvested by sensor $n$ in the preceding time slot $t-1$.

Note that assuming the perfect knowledge of sensors' battery states at the DFC is not practical, because it is extremely costly, if not impossible, for the sensors to synchronize their battery states to the DFC at every time slot.
To reduce the signaling overhead (e.g., energy consumption and channel occupancy), we consider that only if sensor $n$ is activated by the DFC at time slot $t$, it includes the information about its residual energy $\hat e_{n}(t)$ into the update packet.
The channel state information is not available to the DFC when making decisions, which is a common assumption (as in \cite{AoI_Throughput_2019,AoI_EH_Cach_Pappas_2021,EH_AoI_TMC_2021}).
Therefore, if sensor $n$ successfully transmits its update packet in time slot $t$, then the DFC knows its residual energy $\hat e_{n}(t)$. However, due to the randomness of energy arrivals, the sensor's true battery state at the next decision time (i.e., $e_{n}(t+1)$) is still unknown to the DFC. In other words, the sensors' true battery states and availabilities are unobservable to the DFC when making decisions.\footnote{The model can also be extended to the case where the sensor's availability is modeled as an independent two-state time-homogeneous Markov process, as in \cite{Availa_MC_2022}.}
We note that the proposed status update algorithm is executed at the DFC, which is generally powered by the power grid. In this setting, the energy consumption for running the algorithm is not a major issue.

\subsection{AoCI Evolution and Problem Description} \label{Sec:Section 2_AoCI}
The quality of updates generated by one sensor is generally determined by its sensing and computing capacity as well as the spatial relationship between the sensor and its observing target \cite{Dis_TCOM_2021,Distortion_Constraint_2021,Distortion_MSE_2021}. As such, the quality of updates associated with one sensor-CSP pair may be different from the others.
We attribute importance to updates according to their qualities and denote by $D_{n}$ the importance of data updated by sensor $n$ on CSP $k$.
In time slot $t$, the fused information of interest is produced by the DFC, if for all CSPs the qualities of delivered updates are good enough, i.e., the importance threshold is satisfied, as specified in the next paragraph.
In a wireless sensor network, the importance of updates for a sensor can be quantified based on the expected distortion, which can be measured by the mean-squared error (MSE) \cite{Distortion_Constraint_2021,Distortion_MSE_2021}.
Then, for each CSP the importance threshold can be set according to the MSE requirement on the merged data. Similarly, for an edge-assisted video analytics system, the importance associated with each sensor can be defined as the negative of the expected detection accuracy (e.g., the interaction of union (IoU)) regarding its updates \cite{IoU_INFOCOM_2020,IoU_TII_2022}. In this case, the importance threshold for each CSP can be determined by the desired analytical accuracy after fusing the updates delivered from the deployed sensors.
For simplicity, the importance of data updated by each sensor is considered to be fixed by, e.g., ignoring the design of adaptive data processing or resolution adjustment schemes. See Section 6 for extending to the case with adjustable importance.

Let $Y_{n}(t) \in \{0, 1\}, \forall n \in \mathcal N_k, k \in \mathcal K$, indicate whether an update packet from sensor $n$ (monitoring CSP $k$) is successfully delivered in time slot $t$, i.e., $Y_{n}(t) =1$ if it is true, and $Y_{n}(t) =0$ otherwise. To guarantee the quality of the generated integrated information, the importance threshold associated with each CSP $k$ needs to be satisfied. Particularly, for CSP $k$, Let $D_k^0$ and $f_D(({Y_n}(t))_{n \in \mathcal N_k})$ respectively denote the importance threshold and the importance of the aggregated update. The importance condition $f_D(({Y_n}(t))_{n \in \mathcal N_k}) \geq D_k^0$ should be met.
In practice, $f_D(\cdot)$ is determined by various factors, e.g., the physical characteristics of the observed CSP, the distribution of the CSP and sensors, and the fusion mechanism adopted by the DFC. Here, to facilitate the exposition, we consider that
%$f_D(({Y_n}(t))_{n \in \mathcal N_k}) = \frac{1}{\left| {\mathcal N_k(t)} \right|}\sum\nolimits_{n \in \mathcal N_k} D_{n}Y_{n}(t)$,
$f_D(({Y_n}(t))_{n \in \mathcal N_k}) = \sum\nolimits_{n \in \mathcal N_k} D_{n}Y_{n}(t)$.\footnote{Our study can also be extended to cope with the more complex aggregation model incorporating various practical factors.}
We use $\Upsilon_{k}(t) \in \{0, 1\}, \forall k \in \mathcal K$, to indicate whether the importance condition is met, i.e., $\Upsilon_{k}(t) = 1$ if $f_D(({Y_n}(t))_{n \in \mathcal N_k}) \geq D_k^0$, and $\Upsilon_{k}(t) = 0$ otherwise.
A piece of integrated information can be successfully generated only if the importance conditions for all CSPs are satisfied, i.e.,
{\begin{align}  \label{Eq:Succ_Gen_Inte_Info}
\Upsilon(t)  =
\begin{cases}
1, &\textrm{if} \ \sum\nolimits_{k \in \mathcal K} \Upsilon_{k}(t) = K\\
0, & \textrm{otherwise}
\end{cases}
\end{align}
where $\Upsilon(t) =1$ indicates that the integrated information is generated, and $\Upsilon(t) =0$ otherwise.

Let $\Delta(t)$ denote the AoCI at the DFC at the beginning of time slot $t$, which evolves as
\begin{align} \label{Eq:AoI_II_Evolution}
\Delta(t)  =
\begin{cases}
1, &\textrm{if} \ \Upsilon(t-1)=1\\
\Delta(t-1)+ 1, & \textrm{otherwise.}
\end{cases}
\end{align}
Without loss of generality, $\Delta (1)$ is initialized as 0.
It can be readily seen that, due to the correlation among updates, the evolution of AoCI differs from that of the AoI regarding independent sensors.
As demonstrated in Fig. \ref{Fig:Age_variation_Example}, the AoCI would be lowered to the minimum value $1$ only if the integrated information is generated.
In other words, in one time slot only by scheduling ``qualified'' sensors to simultaneously observe all CSPs may reduce the AoCI at the DFC. In this regard, a decision $\mathbf A(t)$ is called valid if
\begin{align}  \label{Eq:Optimal_Nece_Cond}
\sum\nolimits_{k = 1}^K \! \!\mathbf{1} ({f_D(({A_n}(t))_{n \in \mathcal N_k}) \!\geq \!\! D_k^0}) \! = \!\!
\begin{cases}
0, \! &\textrm{if} \ \mathbf A(t) = \mathbf 0\\
K, \! & \textrm{otherwise}
\end{cases}
\end{align}
in which $\mathbf{1}(\cdot)$ represents the indicator function (i.e., 1 if the condition is true, and 0 otherwise) and $f_D(({A_n}(t))_{n \in \mathcal N_k})$ the expected importance of the aggregated update regarding $\mathbf A(t)$.
Since not all the scheduled sensors can be finally activated due to the unobservability of the sensors' true battery states, the following relationship can be established
\begin{align}  \label{Eq:Trans_Resource_Wast}
\sum\nolimits_{n \in \mathcal N_k} \hat A_{n}(t) \leq \sum\nolimits_{n  \in \mathcal N_k} A_{n}(t) \leq  M, \forall k \in \mathcal K.
\end{align}

\begin{figure}[!t]
\centering \includegraphics[width=3.0in]{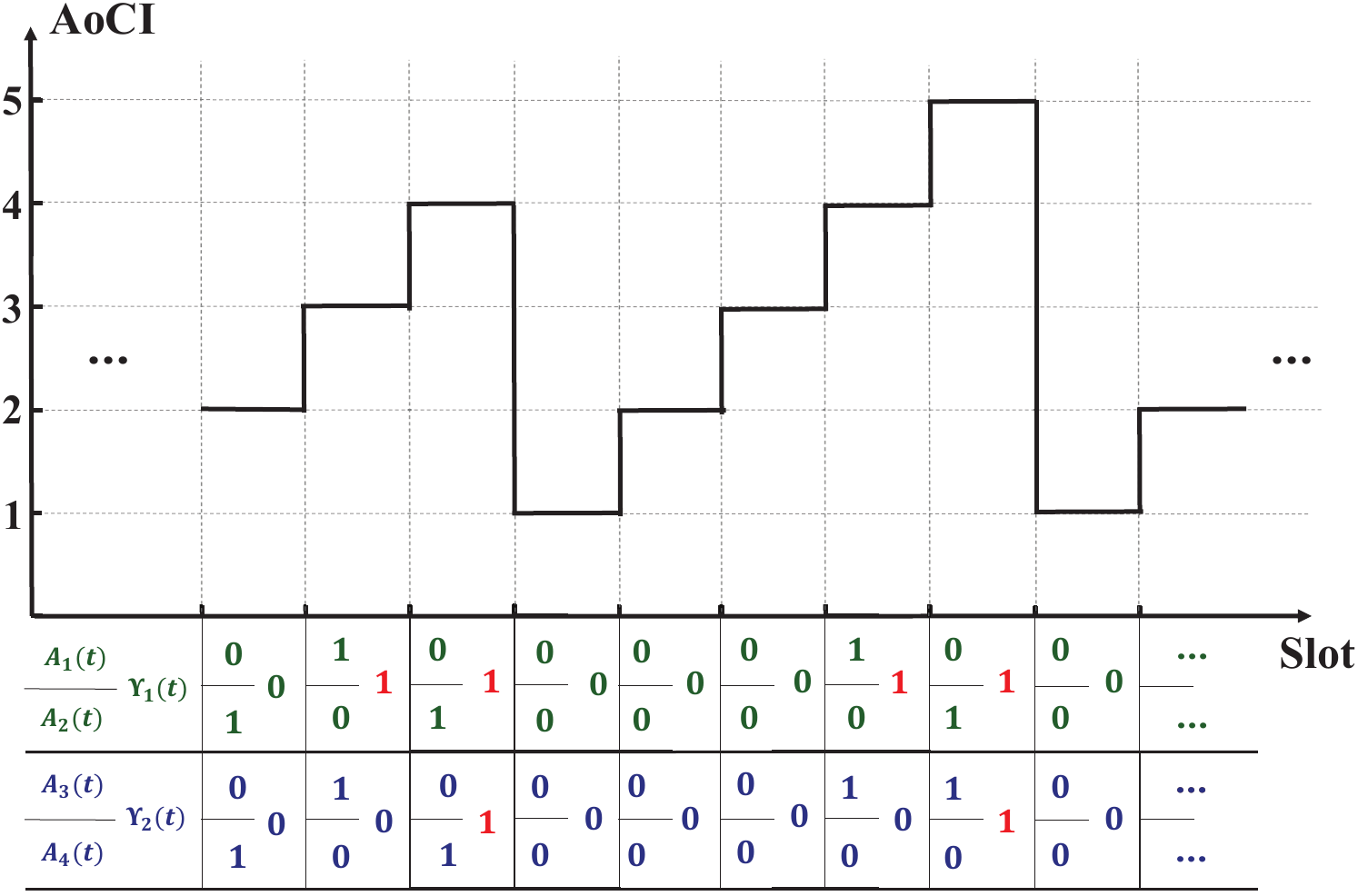}
\centering \caption{An example of the AoCI evolution at the DFC, with $K=2$ CSPs, $N= 4$ sensors, and $M=1$ channel for each set $\mathcal N_k$. For convenience of illustration, here we consider $D_{n} \geq D_k^0,\forall n \in \mathcal N_k, \forall k \in \mathcal K$.} \label{Fig:Age_variation_Example}
\end{figure}

According to (\ref{Eq:AoI_II_Evolution}), to reduce the AoCI, it is crucial to simultaneously activate ``qualified'' sensors to observe all CSPs and successfully deliver their update packets.
Nevertheless, this is non-trivial because of the causality of energy usage, the unknown environmental dynamics (i.e., sensors' transmission failure probabilities and energy arrival rates), and the unobservability of sensors' true battery states.
Essentially, the dynamic status update problem studied in this work is a sequential decision-making problem where the Markov property is preserved, while the system state is partially observable. In light of this, we formulate the problem as a POMDP and develop an RNN-enhanced DRL algorithm to solve it, as elaborated in the following section.
The Lyapunov optimization in queueing networks is also a well-known method for solving the stochastic optimization problem with queue stability constraints, which has been widely adopted to study the dynamic offloading problems in edge computing-enabled wireless networks and IoT systems \cite{LO_IoT_TII_2018,LO_IoT_IoTJ_2019,LO_IoT_TWC_2020,LO_IoT_IoTJ_2021}. However, the Lyapunov framework and the drift-plus-penalty-based algorithms proposed in \cite{LO_IoT_TII_2018,LO_IoT_IoTJ_2019,LO_IoT_TWC_2020,LO_IoT_IoTJ_2021} do not apply to the problem studied in this work. This is because that the fundamental question of this work is how to manage the harvested energy of sensors to send timely status updates. And, due to the information freshness and energy consumption concerns, the generate-at-will model is considered and no buffer queues are introduced for sensors.

\section{POMDP Formulation and Recurrent SAC Status Update Algorithm Design} \label{Section:Algorithm_Des}

\subsection{POMDP Formulation}
We formulate the considered dynamic status update problem as a POMDP, defined by a tuple $\left(\mathbb S, \mathbb O, \mathbb A, U\left({ \cdot , \cdot }\right)\right)$. To facilitate the presentation, we first define several notations. Particularly, for a generic sensor $n$, $\forall n \in \mathcal N_k$, $\forall k \in \mathcal K$, let
\begin{align} \label{Eq:Last_valid_Dur}
    G_{n}(t)&= t -1 - T_{n}(t)
\end{align}
denote the time elapsed since its last successful update to \textit{the beginning of} time slot $t$. Wherein
\begin{align}
T_{n}(t)=\max\{t_0 \left|Y_{n}(t_0)=1,t_0 \leq t-1 \right.\}
\end{align}
represents the delivery time of the latest received update packet that was generated by sensor $n$. Besides, let
\begin{align} \label{Eq:Schedule_times}
X_{n}(t) =
\begin{cases}
0, \! &\textrm{if} \ T_{n}(t) = t-1\\
\sum\nolimits_{l = T_{n}(t)+1}^{t-1}  A_n(l), \! & \textrm{otherwise}
\end{cases}
\end{align}
denote the number of times that sensor $n$ was scheduled during interval $G_{n}(t)$.

Then, we detail the formulated POMDP as follows:

\textbf{1) State space $\mathbb S$}: At time slot $t$, the state of sensor $n$ is defined as $\mathbf S_{n}(t) = \left(G_{n}(t), X_{n}(t), e_{n}(t)\right)$, where the elements $G_{n}(t)$ and $X_{n}(t)$ are respectively specified in (\ref{Eq:Last_valid_Dur}) and (\ref{Eq:Schedule_times}), and $e_{n}(t)$ denotes the sensor's battery state. Furthermore, we define the state of the POMDP as the combination of the states of sensors and the AoCI at the DFC, i.e., $\mathcal S(t) = \{\mathbf S _1(t), \mathbf S _2(t), \ldots, \mathbf S_N(t), \Delta(t)\}$.
Let $\mathbb S$  denote the space of all possible states. To make the state space $\mathbb S$ finite, we set the maximum values of $G_n(t)$, $X_n(t)$ and $\Delta(t)$ as $G_{\max}$, $X_{\max}$, and $\Delta_{\max}$, respectively, which are finite but can be arbitrarily large.

\textbf{2) Observation space $\mathbb O$}: At time slot $t$, let $\mathcal O(t) = \left(\mathbf O _1(t), \mathbf O _2(t), \ldots, \mathbf O_N(t), \Delta(t)\right)$ denote the agent's observation with $\mathbf O_{n}(t) = \left(G_{n}(t), X_{n}(t), \tilde{e}_{n}(t)\right), \forall n \in \mathcal N$. Wherein, $G_{n}(t)$ and $X_{n}(t)$ are respectively defined in (\ref{Eq:Last_valid_Dur}) and (\ref{Eq:Schedule_times}), and $\tilde{e}_{n}(t)$ denotes the observed battery state.
If there was an update successfully delivered from sensor $n$ in time slot $t-1$, then $\tilde{e}_{n}(t)$ is updated as $\tilde{e}_{n}(t)=\hat{e}_{n}(t-1)$; otherwise, $\tilde{e}_{n}(t)$ is set to be a constant value $E_O$ to distinguish it with any valid battery state, i.e., $\tilde{e}_{n}(t) \in \{0, 1, 2, \ldots, E_n\}\bigcup\{E_O\}$, $\forall n \in \mathcal N$.
%The observation $\mathcal O(t)$ is determined by  both the tuple $\left(\mathcal S(t-1), \mathbf A(t-1), \mathcal S(t)\right)$ and the environmental dynamics, which is unknown in advance.
Let $\mathbb O$  denote the space of all possible observations.

\textbf{3) Action space $\mathbb A$}: We denote the space of all valid actions (i.e., decisions satisfying (\ref{equ:Sensing_Action_All}) and (\ref{Eq:Optimal_Nece_Cond})) by $\mathbb A$, i.e.,
\begin{align} \label{Eq:Action_Space}
\mathbb A = \left\{ {\bf{A}}\left| (\ref{equ:Sensing_Action_All}) \& (\ref{Eq:Optimal_Nece_Cond}) \ \textrm{are} \ \textrm{satisfied}\right. \right\}.
\end{align}

\textbf{4) Reward function $U\left({ \cdot , \cdot }\right)$}: At time slot $t$, the reward attained by the agent is defined as
\begin{align} \label{Eq:Utility_function}
U\left({ \mathcal S(t), \mathbf A(t)}\right) = -\Delta(t+1).
\end{align}
The space of all achievable rewards is denoted by $\mathcal U$.

The goal of this work is to find\textit{ a stochastic stationary policy}\footnote{Here, we consider the stochastic policy rather than the deterministic policy, since the former can be beneficial for partially observed problems\cite{POMDP_SP_94,POMDP_SP_15},  as
verified by the simulation results in Section \ref{Sec:Simulation}.} $\pi^*$ that maximizes the long-term discounted accumulative reward \cite{DRQN,Bef_Sta_2008,POMDOP_Survey_2013}, i.e.,
\begin{align} \label{Eq:Optimal_Policy}
\nonumber \pi^* \! &= \mathop {\arg }\limits_ \pi \max \! \mathop {\lim }\limits_{T \to \infty } \!  \mathbb E \!  \big[ \sum\nolimits_{t = 1}^T \gamma^{t-1}\! U\left({ \mathcal S(t), \mathbf A(t)}\right) \left| \mathcal S(1), \mathcal O(1) \right. \! \big] \\
&  \mathop  = \limits^{(a)} \mathop {\arg }\limits_ \pi \min \! \mathop {\lim }\limits_{T \to \infty } \! \mathbb E \!  \big[ \sum\nolimits_{t = 1}^T \gamma^{t-1} \Delta (t) \big]
\end{align}
where action $\mathbf A(t)$ is generated according to the policy $\pi$, the discount factor $\gamma \in [0,1)$  determines the importance of the present reward while ensuring the accumulative reward to be finite, and (a) holds when $\Delta(1)$ is initialized as a constant.

\begin{figure*} [!t]
\begin{align} \label{Eq:Optimal_Policy_MEF}
\pi^* &=\mathop {\arg }\limits_ \pi \max \mathop {\lim }\limits_{T \to \infty }  \mathbb E \big[ \sum\nolimits_{t = 1}^T \gamma^{t-1} \left( U\left({ {\mathcal S}(t), \mathbf A(t)}\right) + \alpha \mathcal H \left(  \pi \left(\cdot  \left| {\mathcal Z}(t) \right.\right) \right)\right) \left| {\mathcal Z}(1) \right. \big]
\end{align}
\hrulefill
\end{figure*}

\begin{figure*} [!t]
\begin{align} \label{Eq:Act_Value_Fun}
Q_{\pi}(\mathcal Z(t),  \mathbf A(t)) = \mathbb E \big[ U\left( {{\cal S}(t),{ \mathbf A}(t)}\right) + \sum\nolimits_{l = 1}^\infty  \gamma^{l}\left(U\left( {{\cal S}(t + l),\pi({\mathcal Z}(t + l))}\right) + \alpha \mathcal H \left(  \pi \left(\cdot  \left| \mathcal Z(t+l) \right.\right) \right)\right)\big]
\end{align}
\hrulefill
\end{figure*}

\begin{figure*} [!t]
\begin{align} \label{Eq:State_Value_Fun}
V_{\pi}(\mathcal Z(t)) = \mathbb E \big[ \sum\nolimits_{l = 0}^\infty  \gamma^{l}\left(U\left( {{\cal S}(t + l),\pi({\mathcal Z}(t + l))}\right) + \alpha \mathcal H \left(  \pi \left(\cdot  \left| \mathcal Z(t+l) \right.\right) \right)\right)\big]
\end{align}
\hrulefill
\end{figure*}

At a typical time slot $t$, the DFC may not be able to obtain the true battery states of all the sensors, but, in principle, can infer this information by going through the entire history of observations $\mathcal Z(t) = (\mathcal O(1), \mathcal O(2), \ldots, \mathcal O(t))$.
Unfortunately, that is generally impractical owing to the huge expenditure on memory storage.
% , while working directly with the entire sequence of observations is hampered by the large memory requirements in practice.
Alternatively, one can use Bayes' theorem to estimate the probability distribution over the state space $\mathbb S$, known as the belief state, which is a sufficient statistic for the history \cite{POMDP_Bel_Sta_1973,POMDOP_Survey_2013,Bef_Sta_2008}.
Nonetheless, the belief update step requires knowledge of the environmental dynamics and, more importantly, may be computationally infeasible when state and action spaces are large.
To address this issue, we incorporate the LSTM network into our algorithm to exploit the historical information in a scalable way.

\subsection{RSS Algorithm Design} \label{Section:Algorithm_Des_Det}
In this part, we first reformulate the original objective (\ref{Eq:Optimal_Policy}) by using the maximum entropy framework and define the corresponding soft action- and state-value functions. Then, conditioned on the maximum entropy formulation (MEF), we proposed a DRL-based dynamic status update algorithm, capitalizing on the SAC and LSTM techniques and incorporating our proposed ADM mechanism.

\subsubsection{\textbf{MEF-based Objective and Soft Value Functions}}
Motivated by the SAC algorithm \cite{SAC_Org_2018}, we aim to train a stochastic actor by augmenting (\ref{Eq:Optimal_Policy}) with the entropy term, and define the MEF-based objective given in (\ref{Eq:Optimal_Policy_MEF}). Wherein, $\mathcal H \left(  \pi (\cdot  \left| \mathcal Z(t) \right.) \right)$ denotes the entropy of the policy conditioned on the history $\mathcal Z(t)$, i.e., $\mathcal H \left(  \pi (\cdot  \left| \mathcal Z(t) \right.) \right)= \mathbb E \big[\! \! - \! \log\pi (\mathbf A  \left| \mathcal Z(t) \right.)\big]$, and $\alpha$ is the temperature parameter determining the relative importance of the entropy versus the reward.
Here, the entropy term is introduced to improve exploration by encouraging diverse behaviors, thereby enhancing the robustness of the learned policy against the model and estimation errors. Additionally, the temperature parameter $\alpha$ is used to adjust the stochasticity of the learned policy, where increasing the value of $\alpha$ leads to a more stochastic policy.
Accordingly, we can define the soft action-value (SA) and soft state-value (SS) functions as shown in (\ref{Eq:Act_Value_Fun}) and (\ref{Eq:State_Value_Fun}), respectively, which satisfy the recursive relationships presented in Theorem 1.

\begin{theorem} \label{The:Consis_Ccond}
For the SA and SS functions respectively defined in (\ref{Eq:Act_Value_Fun}) and (\ref{Eq:State_Value_Fun}), we establish the following recursive relationships
\begin{align} \label{Eq:Action_Value_Fun_Ite}
&\nonumber Q_{\pi}(\mathcal Z(t), \mathbf A(t)) =  \mathbb E \big[ U\left( {{\cal S}(t),{ \mathbf A}(t)}\right) + \gamma V_{\pi}(\mathcal Z(t+1))\big]  \\
&= \sum\nolimits_{U\in\mathcal U, \mathcal S' \in \mathbb S,  \mathcal O' \in \mathbb O} {\mathsf {Pr}_{(\mathcal S(t), \mathbf A(t))}^{(\mathcal S', U, \mathcal O')}} \left(U+\gamma V( \mathcal Z')\right)
\end{align}
with $\mathcal Z' = \{\mathcal Z(t)\}\bigcup\mathcal O'$ and
\begin{align} \label{Eq:State_Value_Fun_Ite}
V_{\pi}(\mathcal Z(t))
 =  \mathbb E_{\mathbf A\sim\pi} \big[Q_{\pi}(\mathcal Z(t),  \mathbf A) - \log \pi (\mathbf A  \left| \mathcal Z(t) \right.)\big].
\end{align}
Wherein, ${\mathsf {Pr}_{(\mathcal S(t), \mathbf A(t))}^{(\mathcal S', U, \mathcal O')}} $ represents the probability that reward $U$ is obtained with the state transitioning from $\mathcal S(t)$ to $\mathcal S'$, if the agent performs action $\mathbf A(t)$ at state $\mathcal S(t)$. Besides, $\mathbf A\sim\pi$ means that action $\mathbf A$ is sampled from policy $\pi$ conditioned on history $\mathcal Z(t)$.
\end{theorem}

\begin{IEEEproof}
The proof is given in Appendix \ref{Pro:Lemm_Consis_Ccond}.
\end{IEEEproof}

Following Theorem~\ref{The:Consis_Ccond}, we can specify the SBR and EKLD (defined later in Section 4.2.2) to learn the SA and SS functions approximated by ANNs.

\subsubsection{\textbf{Architecture of RSS Algorithm}}

\begin{figure*}[!t]
\centering \includegraphics[width=4.5in]{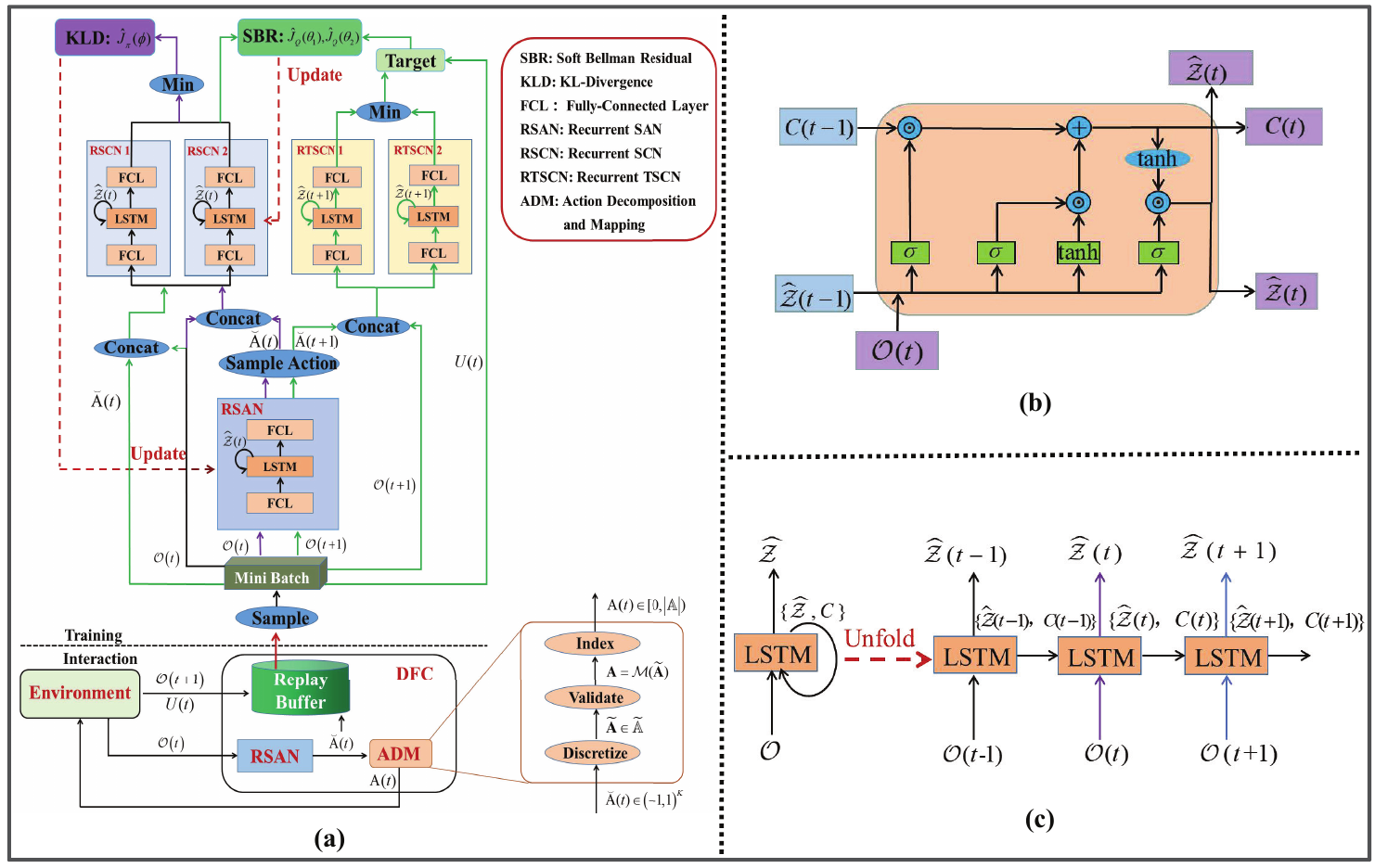}
\centering \caption{Illustrations of the architecture of (a) the RSS algorithm, (b) LSTM cell, and (c) LSTM network unfolded in time.} \label{Fig:Architecture_LSTM}
\end{figure*}

\begin{figure*} [!t]
\begin{align} \label{Eq:Action_Value_Redsidual}
J_Q(\mathbf \theta_j) & = \mathbb E \big[\left(Q_{\mathbf \theta_j}(\mathcal Z(t), \mathbf A(t))-\left(U(\mathcal Z(t), \mathbf A(t))+\gamma\mathbb E\big[V(\mathcal Z(t+1))\big]\right)\right)^2\big] \\ \nonumber
& = \mathbb E \big[\left(Q_{\mathbf \theta_j}(\mathcal Z(t), \mathbf A(t))-\left(U(\mathcal Z(t), \mathbf A(t))+\gamma\mathbb E \big[\mathbb E_{\mathbf A\sim{\pi_{\mathbf {\phi}}}} \big[Q_{\bar{\mathbf \theta}_j}(\mathcal Z(t+1),  \mathbf A) - \log {\pi_{\mathbf {\phi}}} (\mathbf A  \left| \mathcal Z(t+1) \right.)\big]\big]\right)\right)^2\big]
\end{align}
\hrulefill
\end{figure*}

%\begin{figure*} [!t]
%\begin{align} \label{Eq:KL_Div}
%J_\pi(\mathbf {\phi}) = \mathbb E \big[\mathbb E_{\mathbf A\sim{\pi_{\mathbf {\phi}}}} \big[\alpha \log {\pi_{\mathbf {\phi}}}(\mathbf A  \left| \mathcal Z(t) \right.) - Q_{ {\mathbf \theta}}(\mathcal Z(t), \mathbf A) \big]\big]
%\end{align}
%\hrulefill
%\end{figure*}

The architecture of our proposed RSS algorithm is illustrated in Fig. 3 (a), which consists of 5 ANNs, i.e., one recurrent soft actor network (RSAN), 2 recurrent soft critic networks (RSCNs), and 2 target RSCNs (TRSCNs).
The ANNs are constructed based on the SAC \cite{SAC_Org_2018} and LSTM architectures regarding the features of our problem.
The main difference between the RSS and SAC algorithms stems from two aspects: 1) RSS is competent for POMDPs while SAC only applies to MDPs. Particularly, to deal with the partial observability, the LSTM layer is incorporated into the RSCNs and RSAN of RSS, which enables the agent to infer the state information from encountered observations. 2) An ADM mechanism has been developed and embedded in the RSS to circumvent the challenge arising from the large-scale discrete action space, while SAC was developed for the continuous action setting.
%Notably, as shown in the simulation results, by introducing the ADM, both the convergence and stability of our proposed RSS algorithm can be guaranteed even in the scenario with more than $800,000$ valid actions.
In what follows, we first briefly present our designed SAC framework under the assumption that the history $\mathcal{Z}(t)$ can be fully accessed by the agent.
Then, we present the proposed ADM mechanism and RSS algorithm, with which the agent could estimate the current state without directly accessing to $\mathcal{Z}(t)$.

When the history $\mathcal Z(t)$ is accessible, we modify the SAC algorithm \cite{SAC_Org_2018} by changing the input from the current state to $\mathcal Z(t)$. Accordingly, the soft actor network (SAN) is denoted by $\pi_{\mathbf {\phi}}(\mathcal Z(t))$ with $\mathbf {\phi}$ representing its parameters.
To implement the SAC-based architecture for solving our problem, one natural way is to index valid actions with integers from 0 to $\left| {\mathbb A} \right| -1$ and let the SAN's outputs be scalars belonging to $(0, \left| {\mathbb A} \right|)$. Then, valid actions can be obtained by discretizing the outputs (e.g., through rounding down). While significant discretization errors may occur as the action space grows in size, this issue can be effectively addressed by our proposed ADM mechanism, as presented later.
For notation simplicity, we denote the output scale of SAN as the corresponding continuous action before formally introducing the ADM mechanism.

Following \cite{AC_Double_Q_2018}, we introduce two independent ANNs (called soft critic network (SCN) 1 and 2) to approximate the SA function in (\ref{Eq:Act_Value_Fun}) so as to mitigate the bias during training. We respectively parameterize SCN 1 and 2 with $\mathbf \theta_1$ and $\mathbf \theta_2$ and denote them by $Q_{\mathbf \theta_1}(\mathcal Z(t), \mathbf A(t))$ and $Q_{\mathbf \theta_2}(\mathcal Z(t), \mathbf A(t))$.
The SCNs are trained by minimizing their individual SBRs given in (\ref{Eq:Action_Value_Redsidual}), which is derived by applying Theorem \ref{The:Consis_Ccond}. In (\ref{Eq:Action_Value_Redsidual}), $Q_{\bar{\mathbf \theta}_j}(\mathcal Z(t+1),  \mathbf A)$ denotes the target SCN (TSCN) associated with SCN j.
Here, we can update the parameters of SCNs by using stochastic gradient descent (SGD) with backpropagation, and those of TSCNs via the exponential moving average \cite{SAC_Org_2018,DRDPG_2015,DDPG_2016}.

The SAN $\pi_{\mathbf {\phi}}$ is trained by minimizing the EKLD
\begin{align} \label{Eq:KL_Div}
J_\pi(\mathbf {\phi}) \! = \! \mathbb E \big[\mathbb E_{\mathbf A\sim{\pi_{\mathbf {\phi}}}} \big[\alpha \log {\pi_{\mathbf {\phi}}}(\mathbf A  \left| \mathcal Z(t) \right.) \!- \!Q_{ {\mathbf \theta}}(\mathcal Z(t), \mathbf A) \big]\big]
\end{align}
which however cannot be realized by directly using SGD with backpropagation, since the latent variable $\mathbf A(t)$ is sampled according to the SAN $\pi_{\mathbf {\phi}}$ \cite{SAC_Org_2018}.
By applying the reparameterization technique \cite{POMDP_SP_15}, we reparameterize the policy to make $\mathbf A(t)$ differentiable with respect to ${\mathbf {\phi}}$, i.e.,
\begin{align} \label{Eq:Action_RePar}
\mathbf A(t) = f_{\mathbf {\phi}}(\tau,\mathcal Z(t)) = g\left( \mathbf \mu_{\mathbf {\phi}}(\mathcal Z(t)) + \mathbf \sigma_{\mathbf {\phi}}(\mathcal Z(t))\tau\right).
\end{align}
In (\ref{Eq:Action_RePar}), $\mu_{\mathbf {\phi}}(\mathcal Z(t))$ and $\mathbf \sigma_{\mathbf {\phi}}(\mathcal Z(t))$ are the outputs of the SAN, and $\tau$ follows a standard Gaussian distribution with expected value 0 and standard deviation 1.
Besides, $g\left(\cdot\right)$ denotes the generation function that maps unbounded Gaussian samples to $(0, \left| {\mathbb A} \right|)$ by applying a \textit{tanh} activation function and a suitable scaling factor.
By adopting this approach, the gradient $\frac{\partial \mathbf A(t)}{\partial {\mathbf {\phi}}}$ can be backpropagated through the ANNs, facilitating the update of the SAN's parameters.

While valid actions can be obtained by simply discretizing the SAN's output, the discretization error would significantly degrade the algorithm performance when the action space becomes large, as demonstrated in Section \ref{Sec:Simulation}.
This is because, as shown in (\ref{Eq:Action_RePar}), the unbounded Gaussian samples shall be firstly normalized by a \textit{tanh} activation function and then scaled up according to the size of action space $\left| {\mathbb A} \right|$.
As such, the normalized indices of two valid actions become closer as $\left| {\mathbb A} \right|$ increases.
Consequently, it becomes more challenging for DRL algorithms to exactly distinguish them and accurately learn their effects for each state.

\begin{figure*}
\begin{align}  \label{Eq:Action_Subspace}
\mathbb A_k = \{(A_n)_{n\in \mathcal N_k}\left| 0 < \sum\nolimits_{n \in \mathcal N_k} A_n(t) \leq  M \&  {f_D(({A_n}(t))_{n \in \mathcal N_k}) \geq D_k^0} \right.\} \bigcup \{\mathbf 0\}
\end{align}
\hrulefill
\end{figure*}

Unfortunately, the action space of our formulated POMDP grows exponentially with respect to the number of CSPs, $K$, making this simple discretization strategy inefficient. To cope with this issue, we have developed an ADM mechanism by exploiting the structure of our studied problem. Particularly, as presented in (\ref{Eq:Optimal_Nece_Cond}), an update decision is considered valid only if the ``qualified'' sensors are simultaneously scheduled to observe all CSPs. In view of this, we first construct a new action space $\tilde{\mathbb A}$ consisting of $K$ subspaces with smaller sizes, i.e., $\tilde{\mathbb A} = \mathbb A_1 \times \mathbb A_2 \times \cdots \times \mathbb A_K$, regarding the combinations of ``qualified'' sensor(s) associated with each CSP. The expression of each subspace $ \mathbb A_k$ is given in (\ref{Eq:Action_Subspace}), where $\mathbf 0$ denotes the zero vector.
Then, we recast the SAN to make it output $K$-dimensional vectors, coined as primitive actions. A primitive action $\breve{\mathbf A}$ can be expressed as $\breve{\mathbf A} = (\breve A^1, \breve A^2, \ldots, \breve A^K), \forall \breve A^k \in (0, \left| {\mathbb A_k} \right|), \forall k \in \mathcal K$. By discretizing $\breve{\mathbf A}$, we can obtain a $K$-dimensional discrete vector $\tilde{\mathbf A} = (\tilde A^1, \tilde A^2, \ldots, \tilde A^K)$, where $\tilde A^k \in \{0, 1, \ldots, \left| {\mathbb A_k} \right|-1\}, \forall k \in \mathcal K$, denotes the index of an element $\tilde {\mathbf A}_k =(\tilde A_n)_{n\in \mathcal N_k}$ belonging to $\mathbb A_k$.
Hereafter, let $\tilde{\mathbf A}$ denote the corresponding element in $\tilde{\mathbb A}$, which is called as a proto-action.
By implementing this action decomposition, the discretization operation is only performed for each sub-space, which significantly decreases the discretization error since $\left| {\mathbb A_k} \right|\ll\left| {\mathbb A} \right|, \forall k \in \mathcal K$.

However, we can see that the space of actions $\mathbb A$ is a proper subset of $\tilde{\mathbb A}$. Hence, there is no guarantee that the generated proto-action $\tilde{\mathbf A}$ is valid, i.e., $\tilde{\mathbf A} \notin \mathbb A, \exists \tilde{\mathbf A} \in \tilde {\mathbb A}$.
To this end, the following mapping function is proposed
\begin{align} \label{Eq:Mapping_Fun}
\mathbf A = \mathcal M (\tilde{\mathbf A}) =
\begin{cases}
\mathbf 0, &\textrm{if} \ \sum\nolimits_{n \in \mathcal N_k} \tilde {A}_n =0, \exists k \in \mathcal K \\
\tilde{\mathbf A}, & \textrm{otherwise}
\end{cases}
\end{align}
with which each generated proto-action $\tilde{\mathbf A} \in \tilde{\mathbb A}$ can be mapped into the valid action space $\mathbb A$.
The condition in the first case of (\ref{Eq:Mapping_Fun}) indicates that, regarding $\tilde{\mathbf A}$, there is at least one CSP not observed by ``qualified'' sensor(s), for which $\tilde{\mathbf A}$ is mapped to the zero vector, i.e., keeping all sensors silent.

Now, we are ready to complete the RSS algorithm by modifying the constructed ANNs with the LSTM technique. As shown in Fig. \ref{Fig:Architecture_LSTM} (a), we replace the second fully-connected layer (FCL) with the LSTM layer to allow the agent to efficiently learn and estimate the current state without accessing the entire history $\mathcal Z(t)$ upon each time of decision making. For the RSS algorithm, we respectively term the SCN, TSCN and SAN incorporating the LSTM layer as RSCN, RTSCN and RSAN, each of which consists of one LSTM layer and two FCLs. Then, by regarding the output of the LSTM layer as a summarization of the entire history,  we can \textit{equivalently replace the history $\mathcal Z(t)$ in (\ref{Eq:Action_Value_Redsidual})-(\ref{Eq:Action_RePar}) with the observation $\mathcal O(t)$}.
Utilizing the LSTM technique enables the agent to effectively infer the current state based on its observations, thereby making informed decisions. Nevertheless, the limited sequence length and network parameters (memory cells) of the LSTM network inevitably give rise to inference errors.
As a result, this insufficiency can degrade learning performance compared to the scenario where states are entirely observable.

It should be noted that these ANNs are modified by fully considering the features of our formulated POMDP rather than simply following available studies \cite{DRDPG_2015,POMDP_2017,MADRL_TMC_2021}.
Specifically, considering that the dimension of the observation is relatively small in our formulated POMDP, we first expand the observation to a higher dimension by embedding an FCL before feeding it into the LSTM layer, which is different from the ANN architecture devised in \cite{MADRL_TMC_2021}. This design helps to obtain more features from the observation and thus provides more information for the temporal feature extraction, thereby improving the algorithm's performance in terms of both convergence and stability. Besides, for our developed RSAN, only the observation $\mathcal{O}(t)$ is considered as the input, although it is generally recommended to input the action-observation pair $\left(\mathbf{A}(t-1),\mathcal{O}(t)\right)$ to improve the learning performance (see, e.g.,  \cite{DRDPG_2015,POMDP_2017}).
This is because, for our formulated POMDP, the information of $\mathbf{A}(t-1)$ is implicitly contained in two successive observations $\mathcal{O}(t-1)$ and $\mathcal{O}(t)$. In this light, inputting the action-observation pair brings no extra valuable information, but, as per Remark~\ref{Remark:Compu_Complexity}, introduces more input neurons and increases the algorithm's computational complexity.

\subsubsection{\textbf{Details of RSS Algorithm}} \label{Sec:RSS_algorithm_Det}

\begin{algorithm}[!t]
\caption{Recurrent SAC status update (RSS) algorithm.}
\begin{algorithmic}[1] \label{Alg:RSS_algorithm}
\STATE \textbf{Initialization:}
\STATE Initialize the replay buffer $\mathbb D$, temporary buffer $\mathbb D_h$, the parameters of the RSAN, RSCNs and RTSCNs, the start time $W_s$, and training batch size $W_bL$.
\STATE \textbf{Go into a loop:}
\FOR{$episode=1, W$}
\STATE Initialize the observation $\mathcal O(1)$.
\FOR{$t=1, T_w$}
\STATE \textbf{Action selection:} Generate a primitive action $\breve{\mathbf A}(t)$ and its valid counterpart $\mathbf A(t)$ with (\ref{Eq:Action_RePar}) and (\ref{Eq:Mapping_Fun}).
\STATE \textbf{Acting and observing:} Execute the action $\mathbf{A}(t)$, receive the reward $U(t)$, and obtain the new observation $\mathcal{O}(t+1)$.
\STATE \textbf{Updating temporary buffer:} Update the temporary buffer $\mathbb D_h$ as $\mathbb D_h = \mathbb D_h \bigcup \{\breve{\mathbf A}(t), U(t), \mathcal O(t+1)\}$.
\IF {$episode >  W_s$}
\STATE \textbf{Replaying and updating (Training):}
\STATE Sample a mini-batch $W_bL$ of experience tuples $\mathbb F = \{\mathbb F^1, \mathbb F^2, \ldots, \mathbb F^{W_b}\}$ from $\mathbb{D}$.
\STATE Update RSCNs, RSAN, and RTSCNs with (\ref{Eq:Target_Q})-(\ref{Eq:Q_Net_SGD}), (\ref{Eq:Policy_Net_SGD})-(\ref{Eq:Q_WT_Pi_Old}), and (\ref{Eq:Target_Q_Update}), respectively
\ENDIF
\ENDFOR
\STATE \textbf{Refreshing replay buffer:} Transfer the experience tuples in $\mathbb D_h$ into $\mathbb{D}$, and clear $\mathbb D_h$.
\ENDFOR
\STATE \textbf{Output:} The trained RSAN.
\label{Alg1:Iteration_End}
\end{algorithmic}
\end{algorithm}

\begin{figure*} [!t]
\begin{align} \label{Eq:Target_Q}
\bar Q_{t_w+l} = U_{t_w+l} + \gamma \left(\mathop {\min }\limits_{j \in \{1,2\}} Q_{\bar {\mathbf \theta}_j}(\mathcal O_{t_w+l+1}, \breve { \mathbf A}_{t_w+l+1}) - \alpha \log {\pi _\phi }( \breve {\mathbf A}_{t_w+l+1}|\mathcal O_{t_w+l+1})\right)
\end{align}
\hrulefill
\end{figure*}

The details of our proposed RSS algorithm are presented in Algorithm \ref{Alg:RSS_algorithm}.\footnote{Code is at: https://github.com/CXU-NWAFU/RSS\_algorithm.}
Firstly, the experience replay buffer $\mathbb D$ and temporary buffer $\mathbb D_h$ are cleared, the parameters of the RSAN and RSCNs, i.e., $\mathbf {\phi}$, $\mathbf \theta_1$ and $\mathbf \theta_2$, are randomly initialized, and the parameters of the RTSCNs are respectively set as $\bar {\mathbf \theta}_1 = \mathbf\theta_1$ and $\bar {\mathbf\theta}_2 = \mathbf\theta_2$. The learning process is divided into $W$ episodes, each of which comprises $T_w$ time slots.
When the initialization is completed, the algorithm goes into a loop. At the beginning of each iteration (episode), the observation $\mathcal O(1)$ is initialized by setting $\tilde{e}_{n}(t) = E_n$, $\forall n \in \mathcal N$, and other elements as 0.
Then, in time slot $t$, a primitive action $\breve{\mathbf A}(t)$ and its valid counterpart $\mathbf A(t)$ are generated with (\ref{Eq:Action_RePar}) and (\ref{Eq:Mapping_Fun}).
%\footnote{The action generated by the RSAN and that belonging to the action space $\mathbb A$ are called as the primitive action and valid action, respectively.}
After the valid action $\mathbf A(t)$ is executed, a certain reward $U(t)$ and a new observation $\mathcal O(t+1)$ are obtained by the agent. And, the temporary buffer is updated as $\mathbb D_h = \mathbb D_h \bigcup \{\breve{\mathbf A}(t), U(t), \mathcal O(t+1)\}$.
Here, by buffering the primitive action instead of the valid action, the gradients of our constructed ANNs can be backpropagated in the same way as in the SAC algorithm. However, since a set of primitive actions may be mapped to the same valid action, similar long-term effects of them (in a given state) may be learned.
In $\mathbb D_h$, the successive 4 experience elements associated with a generic time slot $t$ are coined as an experience tuple and briefly denoted by
$\mathcal F_t =\{\mathcal O_{t}, \breve{\mathbf A}_{t}, U_{t}, \mathcal O_{t+1}\}$.
At the end of each episode, all experience tuples in buffer $\mathbb D_h$ are transferred into buffer $\mathbb D$ for future replay.

After $W_s$ episodes are completed, the training process begins. To be specific, we first randomly select $W_b$ episodes with the experiences already stored in $\mathbb D$ and then, randomly sample $L$ successive experience tuples from each selected episode.
For one selected episode $w$, we denote the time slot regarding the first sampled experience tuple by $t_w$ and the set of sampled tuples by $\mathbb F^w = \{\mathcal F_{t_w}, \mathcal F_{t_w+1}, \ldots, \mathcal F_{t_{w+L-1}}\}$, with $\mathcal F_{t_w} =\{\mathcal O_{t_w}, \breve{\mathbf A}_{t_w}, U_{t_w}, \mathcal O_{t_w+1}\}$.
For each sampled tuple $\mathcal F_{t_w+l}$, we compute the target SA $\bar Q_{t_w+l}$ by resorting to (\ref{Eq:Target_Q}). Wherein, $\breve {\mathbf A}_{t_w+l+1}$ denotes the primitive action generated from the RSAN $\pi_{\mathbf {\phi}}$ with the observation $\mathcal O_{t_w+l+1}$ as the input. Furthermore, the SBR (\ref{Eq:Action_Value_Redsidual}) for each RSCN $j$ ($\forall j \in \{1, 2\}$) can be approximated as follows:
\begin{align} \label{Eq:Action_Value_Redsidual_Approx}
&\hat J_Q(\mathbf \theta_j) \\ \nonumber
&= \frac{1}{W_b L}\sum\nolimits_{w = 1}^{W_b}\sum\nolimits_{l = 1}^L\left(Q_{\mathbf \theta_j}(\mathcal O_{t_w+l}, \breve{{\mathbf A}}_{t_w+l})-\bar Q_{t_w+l}\right)^2.
\end{align}
And, the parameters of RSCN $j$ are updated by SGD, i.e.,
\begin{align} \label{Eq:Q_Net_SGD}
\mathbf \theta_j = \mathbf \theta_j - \eta_j \nabla_{\theta_j} \hat J_Q(\mathbf \theta_j)
\end{align}
with $\eta_j$ denoting the learning rate.

Similarly, the RSAN's parameters $\phi$ can be updated as
\begin{align} \label{Eq:Policy_Net_SGD}
\mathbf \phi = \mathbf \phi - \eta_0 \nabla_{\mathbf \phi} \hat J_{\pi}(\mathbf \phi)
\end{align}
in which $\eta_0$ denotes the learning rate, and $\hat J_{\pi}(\mathbf \phi)$ is the approximation of the EKLD in (\ref{Eq:KL_Div}), i.e.,
\begin{align} \label{Eq:KL_Div_Approx}
&\hat J_{\pi}(\mathbf \phi) \\ \nonumber
&= \frac{1}{W_b L}\sum\nolimits_{w = 1}^{W_b}\sum\nolimits_{l = 1}^L\alpha \log {\pi_{\mathbf {\phi}}}(\breve{\mathbf A}_{t_w +l}^{\mathbf {\phi}} | \mathcal O_{t_w+l}) - \hat Q_{t_w+l}.
\end{align}
In (\ref{Eq:KL_Div_Approx}), $\breve{\mathbf A}_{t_w +l}^{\mathbf {\phi}}$ denotes the primitive action generated by the RSAN $\pi_{\mathbf {\phi}}$ with $\mathcal O_{t_w+l}$ as the input, and the last term on the right-hand side, i.e., $\hat Q_{t_w+l}$, can be expressed as
\begin{align} \label{Eq:Q_WT_Pi_Old}
\hat Q_{t_w+l} = \mathop {\min }\limits_{j \in \{1,2\}} Q_{{\mathbf \theta}_j}( \mathcal O_{t_w+l}, \breve{\mathbf A}_{t_w +l}^{\mathbf {\phi}}).
\end{align}
At the end of each training round, the parameters of RTSCNs are updated via the exponential moving average, i.e.,
\begin{align} \label{Eq:Target_Q_Update}
\bar {\mathbf \theta}_j = \tau \mathbf \theta_j + (1-\tau) \bar {\mathbf \theta}_j, \forall j \in \{1,2\}
\end{align}
where $0<\tau \ll 1$ is a constant utilized to ensure that the target networks change slowly while effectively keeping track of the RSCNs being trained.

\begin{remark} \label{Remark:Compu_Complexity}
\textbf{Computational Complexity.} For the LSTM network comprises an input layer with $q_I$ neurons, a recurrent LSTM layer with $q_C$ neurons (memory cells), and an output layer with $q_O$ neurons, the computational complexity of a training step is $\mathcal O(q_C(4q_I+4q_C+q_O+3))$ \cite{LSTM_Complexity}. Besides, for a fully-connected ANN with $q_{I}$ input and $q_{O}$ output neurons, the training computational complexity is $\mathcal O(q_{I}q_{O})$ \cite{Wu_2016_CVPR}. For our developed RSS algorithm, there are one RSAN and two RSCNs to be trained, each of which consists of one input, one output, and two hidden layers (i.e., one FCL and one LSTM layer). To this end, the training computational complexity of the RSAN can be expressed as
\begin{align}
\mathcal O_{A}=\mathcal O\left(( \mathbb O)_Dq_{A}^{F}+q_{A}^{C}(4q_{A}^{F}+4q_{A}^{C}+q_{A}^{O}+3)\right)
\end{align}
where $( \mathbb O)_D = 3N+1$, $q_{A}^{F}$, $q_{A}^{C}$ and $q_{A}^{O}$ denote the number of neurons in the input layer, hidden FCL, hidden LSTM layer, and output layer, respectively. Similarly, for RSCN $j$, $\forall j \in \{1, 2\}$, the training computational complexity can be expressed as
\begin{align}
\mathcal O_{C_j} \!=\!\mathcal O\!\left(( (\mathbb O)_D\!+\!K)q_{C_j}^{F}+q_{C_j}^{C}(4q_{A}^{F}\!+\!4q_{C_j}^{C}\!+\!q_{C_j}^{O}\!+\!3)\right)
\end{align}
where $q_{C_j}^{F}$, $q_{C_j}^{C}$, and $q_{C_j}^{O}=1$ respectively denote the numbers of neurons in the hidden FCL, hidden LSTM layer, and output layer.

In view of this, the training computational complexity of our proposed RSS algorithm can be expressed as $\mathcal O_{A}+\mathcal O_{C_1}+\mathcal O_{C_2}$, which incorporates the training of the RSAN and two RSCNs. Similarly, the computational complexity for one inference (i.e., action generation) can be expressed as $\mathcal O_{A}$, since only the RSAN is utilized during the decision-making process. Actually, as demonstrated in Section \ref{Sec:Simulation}, the computational complexity of our proposed algorithm for one inference is very low, thereby allowing the system to make real-time decisions on status update in practice\footnote{
Note that for the real-time application in which the requirement of timeliness is on the order of $\mu$s, e.g., machine tools in factory automation \cite{Real_Time_App_2017}, it is not recommended to transmit status updates via wireless links and our proposed algorithm does not apply.
}. For instance, when there are $28$ sensors and $823544$ valid actions, the inference latency is merely about $0.44$ ms.
\end{remark}

We note that although a rigorous convergence analysis of our proposed algorithm is not available because this is still an open problem for DRL algorithms, the simulation results in the next section demonstrate that the convergence of our proposed SAC algorithm can be well achieved even in the scenario with more than $800,000$ valid actions.

\section{Simulation Results}  \label{Sec:Simulation}
%We conduct simulations to verify the effectiveness of our proposed RSS algorithm. Specifically, we first present the simulation setting, and then evaluate the effectiveness of our proposed RSS algorithm by comparing it with several baseline DRL algorithms.

\subsection{Simulation Setting} \label{Sec:Simulation_Setting}
Unless otherwise specified, the default setting is as follows. We consider the IoT network with one DFC and $K$ CSPs, each of which can be observed by $4$ sensors.
There are $M=2$ available orthogonal channels for each sensor set.
To consider the networks with different scales, we vary the number of CSPs $K$ from 3 to 7, i.e., the number of sensors $N$ varies from 12 to 28.
Due to the heterogeneity of sensors in each set $\mathcal N_k$, their transmission failure probabilities are considered to be different \cite{AoI_Throughput_2019,AoI_Pef_SCI,Trans_Succ_Pro_TMC_2021}.
Particularly, the sensors' transmission failure probabilities are set as $\{0.05, 0.10, 0.15, 0.20\}$, and the importance of their generated status updates are set as $\{0.4, 0.6, 0.8, 1.0\}$ with the importance threshold being $1.0$.
This setting indicates that an update packet with a higher quality would be successfully transmitted with a lower probability. For instance, the high definition image has a better quality and a larger data size than the standard definition image, leading to a higher transmission failure probability. For each sensor, its EH probability and battery capacity are respectively set to $0.2$ and 20. Meanwhile, we set $G_{\max} = X_{\max} = 2\Delta_{\max} = 4NK$,  $\gamma = 0.99$, and $T_w=10^3$ time slots. As in \cite{EH_Pro_JSAC_2016,EH_AoI_TMC_2021}, the duration of a time slot is not specified, since it depends on the adopted IoT device and system protocol.

\begin{table}[!t]
\renewcommand{\arraystretch}{1.0}
\caption{Setting of hyperparameters}
\label{Tab:Hyperparameters}
\centering
\begin{tabular}{ |l | l | }  \hline
\textbf{Hyperparameter and description} & \textbf{{Setting}} \\ \hline
%\midrule
Learning rates, $\eta_0, \eta_1, \eta_2$                                      & $5\times10^{-4}$                                                 \\ \hline
Target update rate, $\tau$                                                                            & $1\times10^{-3}$                                           \\ \hline
%The equivalent processing rate for different data, $\frac {\gamma _j}{r}$, $\forall j \in \{1, 2, 3\}$              & 50 Mbits/s                                        \\
Replay buffer size, $\left| { \mathbb D} \right|$                                                         & $10^3$                                               \\ \hline
Replay start time, $W_s$                                                                                 & $2\times10^2$                                        \\ \hline
Mini-batch size, $W_b$                                                                            & $10$                                           \\ \hline
Sequence length, $L$                                                                                      & 50 \\ \hline
Temperature parameter, $\alpha$                                                                   & $2\times10^{-2}$     \\ \hline
%Optimizer                                                                                                 & Adam
Optimizer                                                                                                 & RMSprop\\ \hline
%Weights initializer                                                                                       & He
Weights initializer                                                                                       & He
\\
%\toprule
\hline
\end{tabular}
\end{table}

\begin{figure*} [!t]
\centering
\subfigure[] {\leavevmode \epsfxsize=2.36in  \epsfbox{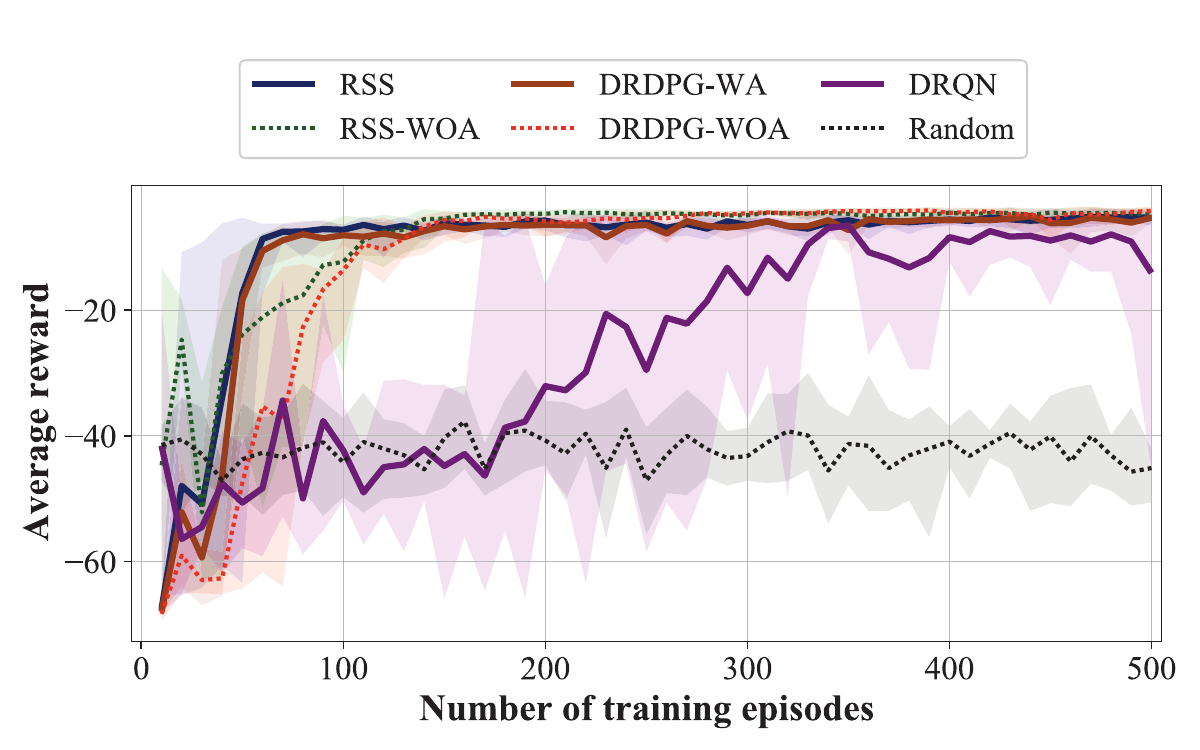}}
\subfigure[] {\leavevmode \epsfxsize=2.36in  \epsfbox{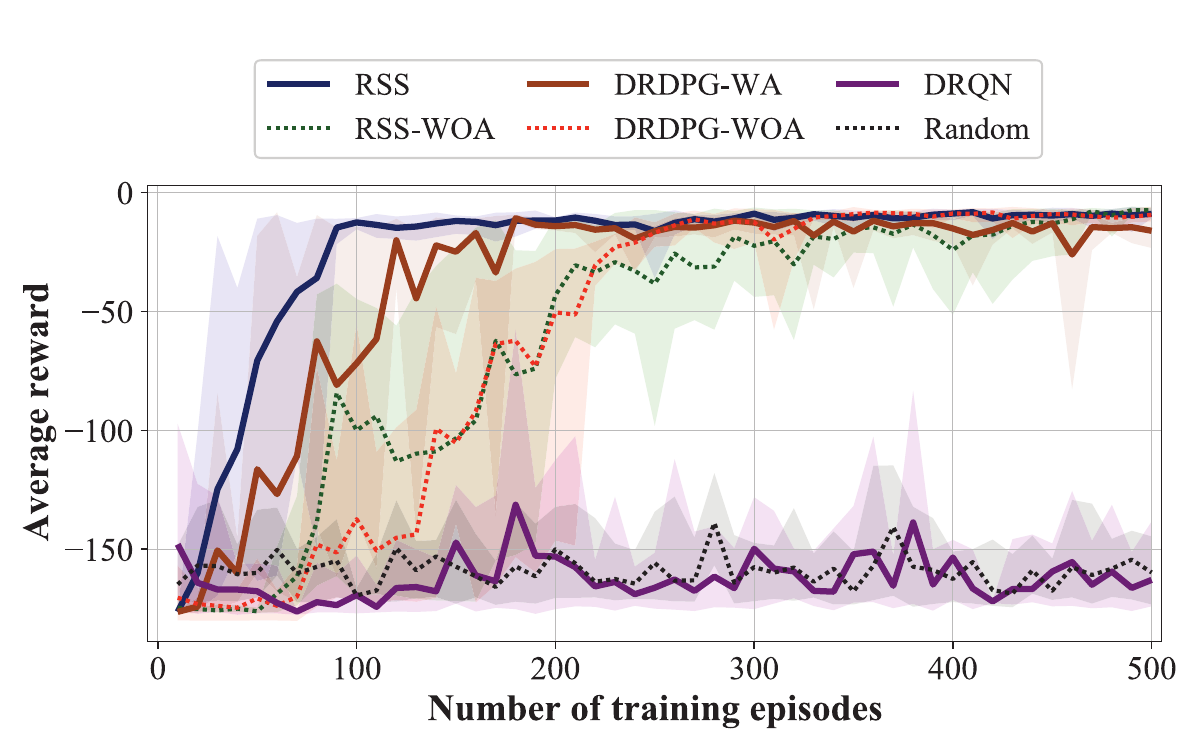}}
\subfigure[] {\leavevmode \epsfxsize=2.36in  \epsfbox{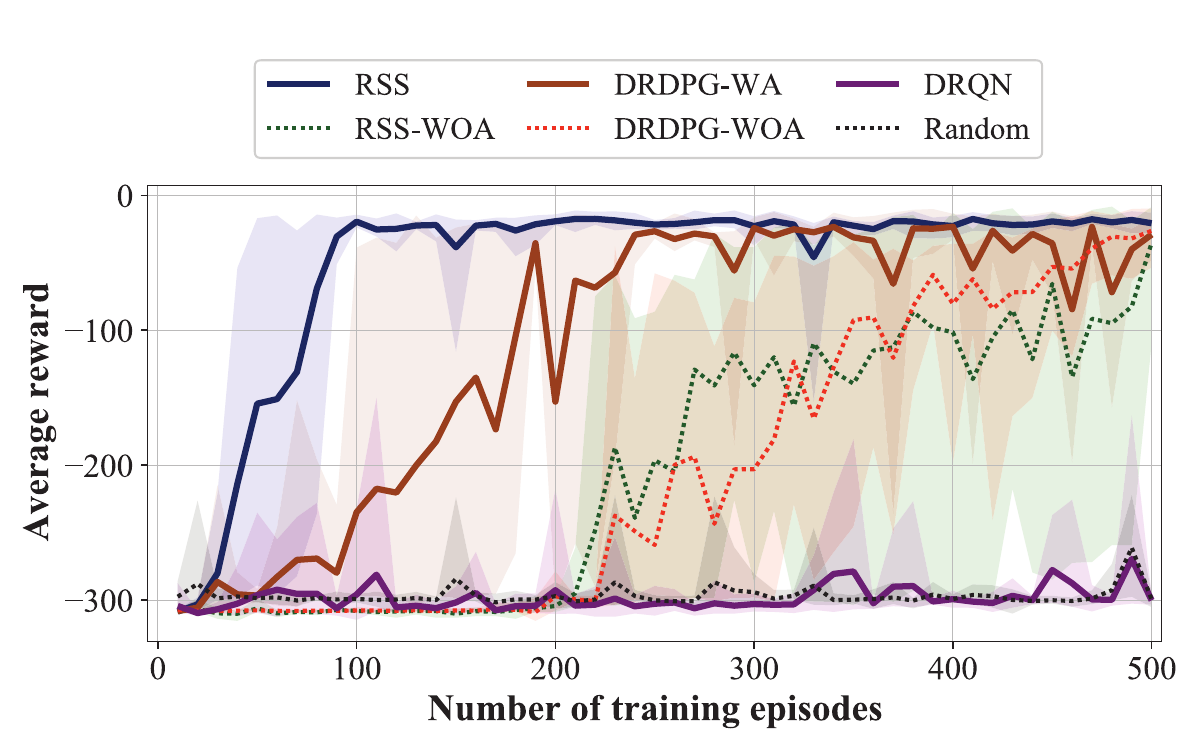}}
\centering \caption{Convergence comparison in IoT networks with: (a) $K=3$ ; (b) $K=5$; (c) $K=7$ CSPs.}
\label{Fig:Convergence}
\end{figure*}

To evaluate the effectiveness of our proposed RSS algorithm, we compare its performance with the random policy and four DRL algorithms, i.e., DRQN \cite{DRQN}, DRDPG-WA (DRDPG \cite{DRDPG_2015} with ADM), DRDPG-WOA (DRDPG without ADM), and RSS-WOA (RSS without ADM).
In recent work \cite{MADRL_TMC_2021}, a multi-agent DRL algorithm was proposed to efficiently navigate a group of unmanned vehicles for data collection and battery charging, by combining the multi-agent DDPG (MADDPG) algorithm \cite{MADDPG_NIPS} with the Ape-X \cite{Ape_X_2018} and LSTM architectures. When implementing this algorithm to our problem, the algorithm boils down to the DRDPG enhanced by the Ape-X architecture.
It is worth emphasizing that with the aid of the Ape-X architecture, any off-policy DRL algorithm can be implemented in a distributed setting with multiple samplers.
However, this is achieved at the expense of additional computing resources, since samplers need to run on different CPUs/CPU cores to perform distributed experience generation \cite{Ape_X_2018}.
Although it is also possible to accelerate our proposed RSS algorithm with the Ape-X architecture, this aspect is beyond the scope of this paper.
In the following, the performance is compared in terms of the achieved average reward, defined as per (\ref{Eq:Utility_function}), i.e., the negative AoCI at the end of each time slot.

For our proposed RSS algorithm, all ANNs (i.e., the RSAN, RSCNs, and RTSCNs) consist of one input, one output, and two hidden layers, as presented in Fig. \ref{Fig:Architecture_LSTM} (a).
The first hidden layer is an FCL with 128 neurons, and the second hidden layer is an LSTM layer also with 128 neurons. For RSAN, its input is the observation and there are $2K$ neurons used for the output layer to generate the primitive action. Meanwhile, for RSCNs and RTSCNs, the input is an observation-primitive action pair, and only one neuron is used for the output layer to estimate the SA. For a fair comparison, the ANNs utilized by the baseline DRL algorithms also consist of two hidden layers (one FCL and one LSTM layer), each of which has $128$ neurons.
%Based on the experiments in \cite{SAC_Org_2018}, we set the temperature parameter $\alpha$ to 0.001.
To achieve exploration, DRQN adopts the $\varepsilon$-greedy policy with $\varepsilon$ annealing linearly from 1.0 to 0.01 \cite{Rainbow_DQN_Para}, while DRDPG utilizes a Gaussian policy with exploration noise following Gaussian distribution $N(0, 0.1)$ \cite{AC_Double_Q_2018}. For all DRL algorithms, the inputs are normalized. The other hyperparameters are provided in Table \ref{Tab:Hyperparameters}. All simulations are conducted by using a single NVIDIA GPU of GeForce RTX 2080 Ti, and the CPU is Intel(R) Core(TM) i9-9900K@3.60GHz with 64 GB RAM.

%The architecture of the two critic networks are totally same, as well as the corresponding target networks. Between each layer for the actor and critics, the activation function of each neuron is Rectified Linear Unit (ReLU), which is equal to itself when it is greater than zero, otherwise equals zero. In particularly the activation function of the output layer of the actor is tanh. The actor receives state as input to the first layer while the critics receives both state and action which normalized by the maximum value of each element.

%\begin{itemize} \setcounter{enumi}{0}
%\item DRQN: Deep R-network based dynamic status update algorithm, a combination of the standard DQN and R-learning, whose pseudo-code is given in Appendix \ref{Append:DR_DDU}.
%\item DDQ-DSU: Dueling DQN based dynamic status update algorithm, where the dueling architecture is introduced into the standard DQN \cite{Dueling_DQN_Org}. Its original aim is to maximize the discounted long-term cumulative reward instead of the average one.
%\item DQ-DSU: DQN based dynamic status update algorithm, where the standard DQN \cite{DQN_Nature_Letter} is utilized. Its original aim is to maximize the discounted long-term cumulative reward instead of the average one.
%\item Greedy policy: With this policy, the ECN will ask sensors with the stalest $M$ cached data packets to update their current status at the beginning of each time step.
%\item Random policy: With this policy, the ECN will randomly choose an available action (i.e., $\sum\nolimits_{k = 1}^K a_k(t) \leq M$) at the beginning of each time step.
%\end{itemize}

\subsection{Convergence Comparison}

We first investigate the convergence performance of our proposed RSS algorithm. To test the convergence of DRL algorithms, we extract the trained RSAN to make decisions after each training episode, during which $\varepsilon = 0.05$ is adopted for the DRQN algorithm \cite{Rainbow_DQN_Para}.
The average reward is calculated by interacting with the environment for one episode, and the training ends after 500 episodes.
All simulation results are obtained by averaging over six independent runs with different seeds, while the same seed is adopted for all algorithms in one run for fair comparison.
The convergence performance comparison is presented in Fig. \ref{Fig:Convergence}, where the darker (solid or dotted) line shows the average value over runs, and the shaded area is obtained by filling the interval between the maximum and minimum values.
Notably, with $K = 3, 5$, and $7$ CSPs, there are $344$, $16,808$, and $823,544$ valid actions, respectively.
Alternatively, one can keep $K$ constant while increasing the number of sensors associated with each CSP to expand the action space. Nevertheless, the setting adopted in this work requires more CSPs to be simultaneously monitored, which makes the learning process more challenging and interesting.

\begin{figure*} [!t]
\centering	
\subfigure[] {\leavevmode \epsfxsize=2.30in  \epsfbox{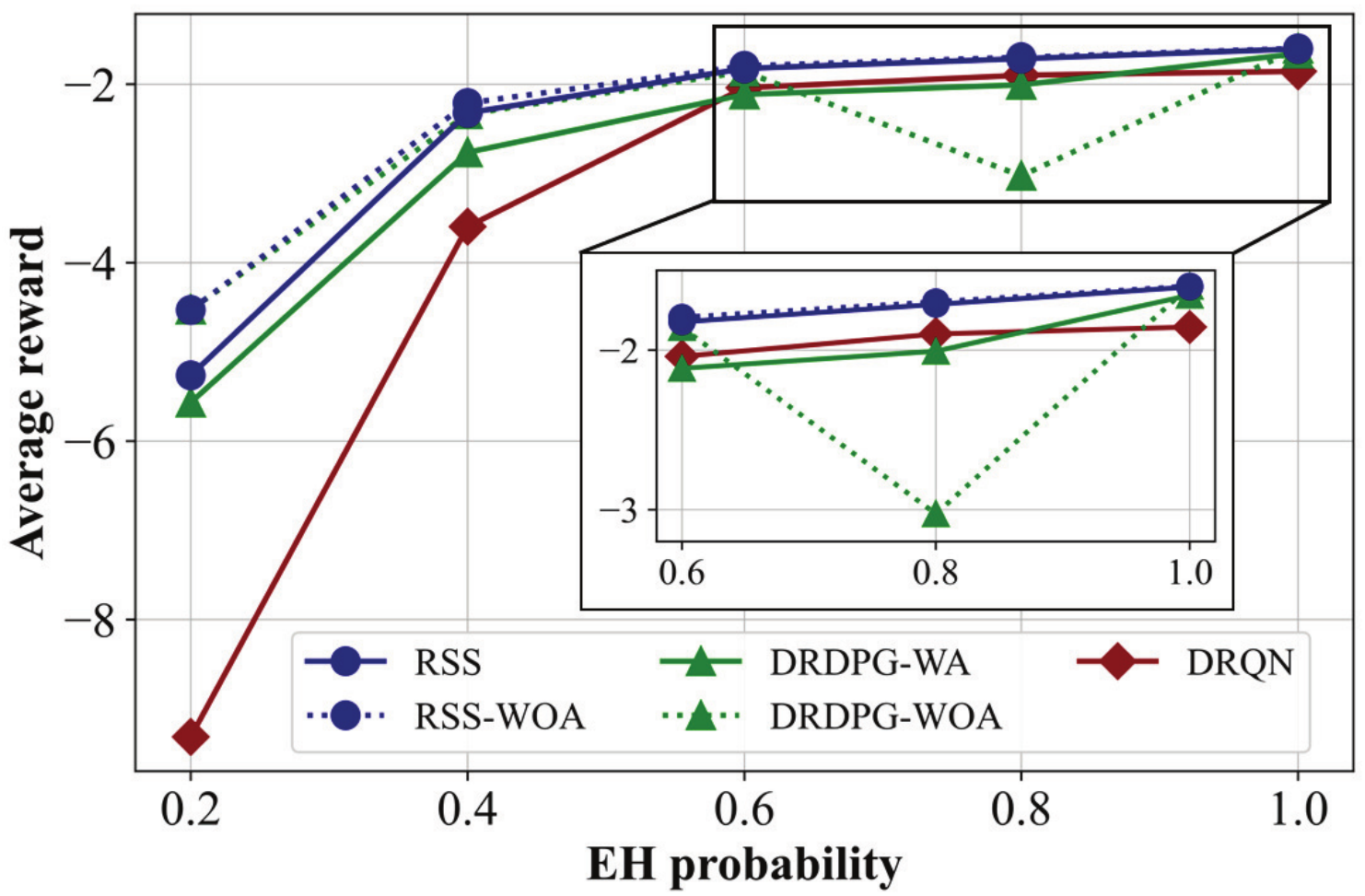}}
\subfigure[] {\leavevmode \epsfxsize=2.30in  \epsfbox{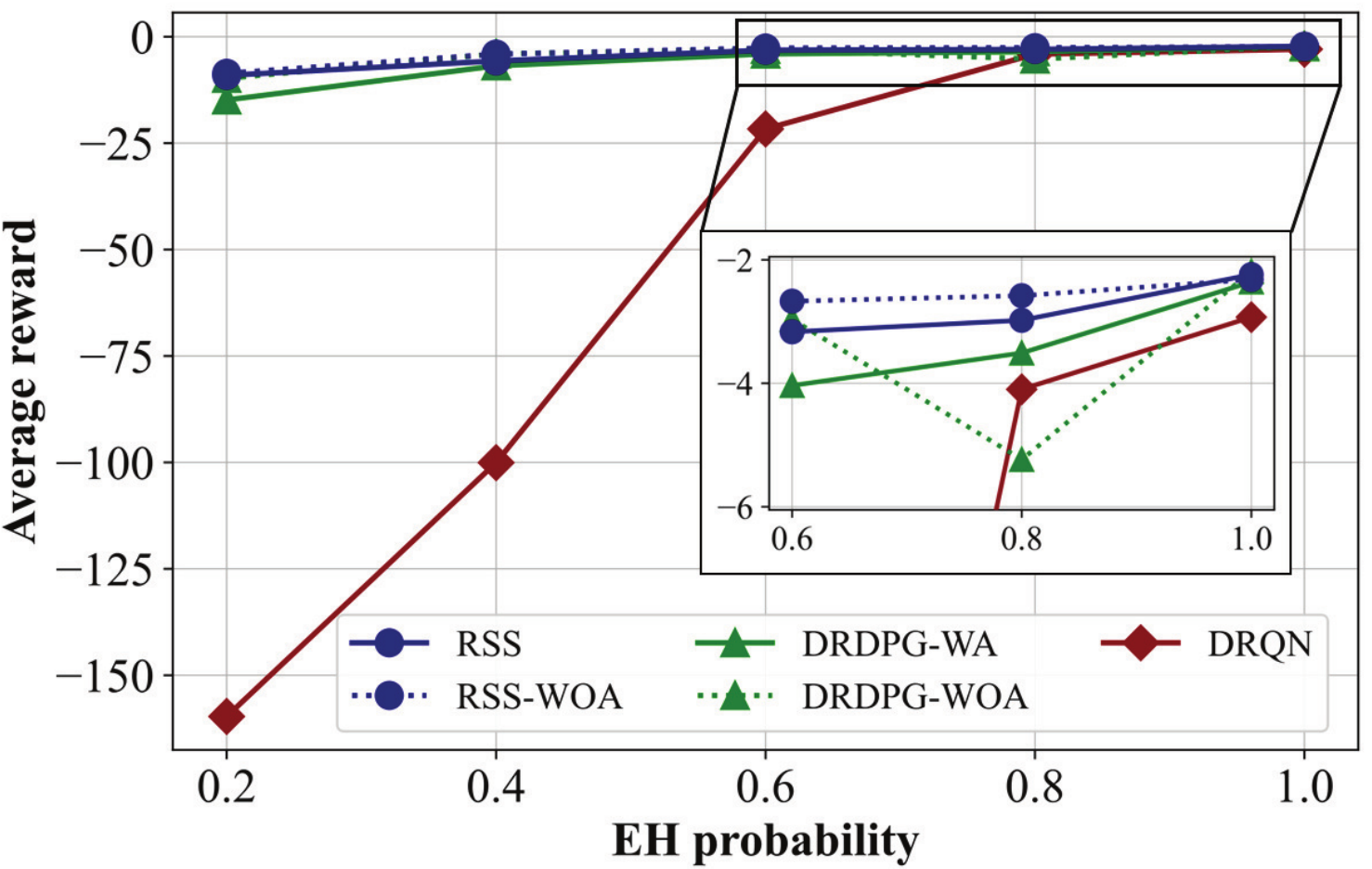}}
\subfigure[] {\leavevmode \epsfxsize=2.30in  \epsfbox{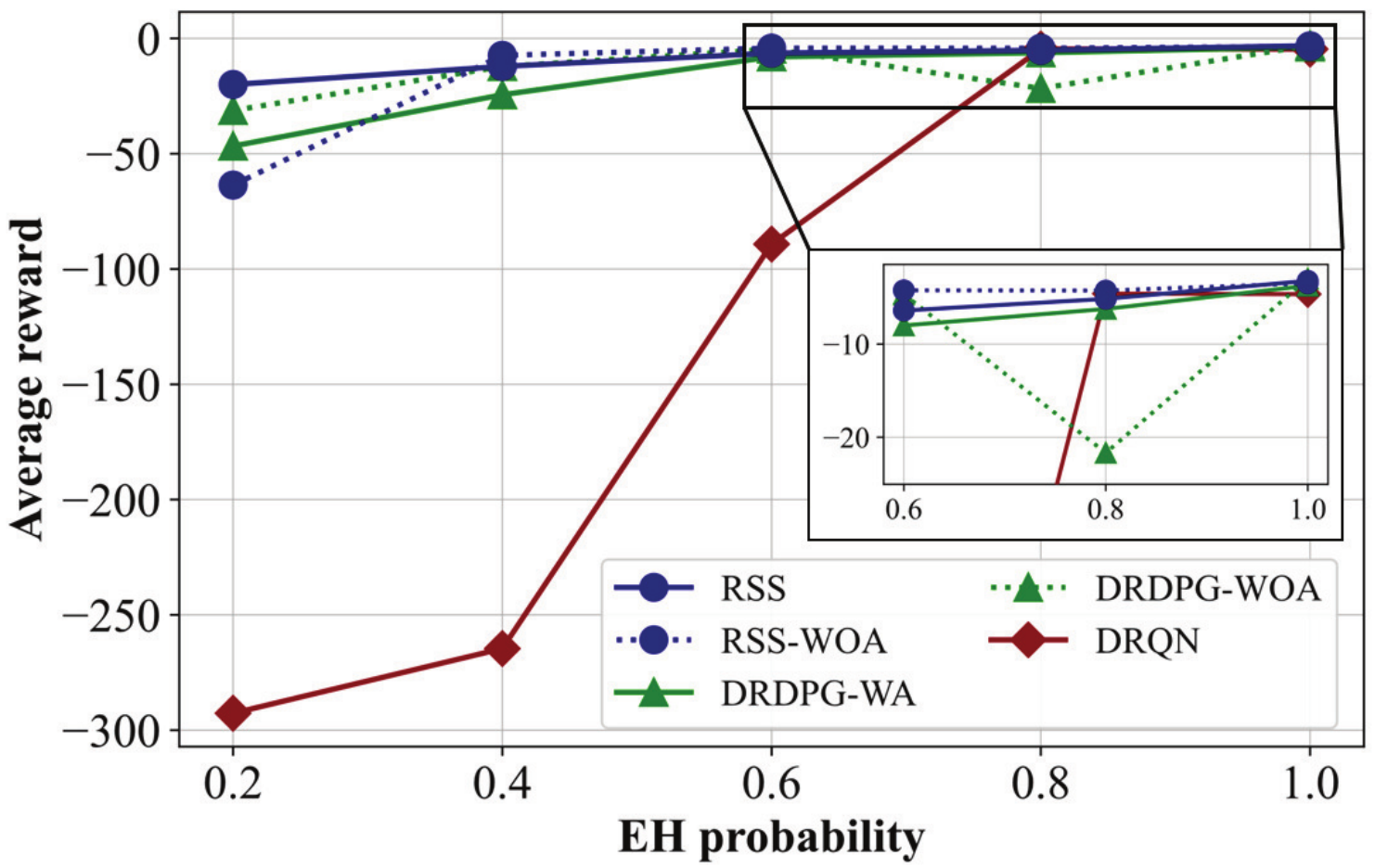}}
\centering \caption{Performance comparison in the IoT network with: (a) $K = 3$ ; (b) $K=5$; (c) $K=7$ CSPs, where the EH probability varies from 0.2 to 1.0.}
\label{Fig:change_ehp_3_5_7CSPs}
\end{figure*}

Several observations can be made from Fig. \ref{Fig:Convergence}.
First, the RSS algorithm's convergence is guaranteed even for the case with $823,544$ valid actions (i.e., $K = 7$ CSPs), which however cannot be well dealt with by the baseline DRL algorithms. Particularly, for DRDPG-WOA and RSS-WOA algorithms, the convergence can only be seen in the cases with $K = 3$ and $K = 5$ CSPs. This is because the larger the action space, the greater the learning error will be introduced by the simple discretization operation, as discussed in Section 4.2.2.
Second, incorporating the ADM mechanism into the DRL algorithms significantly accelerates the convergence.
For instance, when $K=5$, RSS and RSS-WOA converge after about 90 and 330 training episodes, respectively.
Moreover, after the convergence is attained, the performance of RSS and RSS-WOA algorithms are respectively superior to their DDPG-based counterparts, i.e., DRDPG-WA and DRDPG-WOA, in terms of stability and achieved average reward. This improvement is mainly attributed to the fact that, compared with the deterministic policy, the stochastic policy is more beneficial for partially observed problems \cite{POMDP_SP_94,POMDP_SP_15}.
To further demonstrate this, we present the mean and standard deviation (SD) of the achieved average reward during the last 50 evaluations for the Random policy and DRL-based algorithms in Table~\ref{Tab:Mean_Std_Conv}, in which the best results are marked in bold.
It can be seen that RSS outperforms RSS-WOA in the large-scale scenario (i.e., $K = 7$), while being slightly worse than RSS-WOA in the small-scale scenario (e.g., $K=3$ or $K=5$).

\begin{table}[!t]
\centering
\caption{Mean and SD of the average reward achieved by different algorithms during the last 50 evaluations.}
\label{Tab:Mean_Std_Conv}
\begin{tabular}{|c|c|c|c|c|c|c|}
\hline
\multirow{2}{*}{\textbf{Algorithm}} & \multicolumn{3}{c|}{\textbf{Mean}}                & \multicolumn{3}{c|}{\textbf{Standard Deviation}} \\ \cline{2-7}
            & \textit{K}=3   & \textit{K}=5   & \textit{K}=7    & \textit{K}=3  & \textit{K}=5  & \textit{K}=7  \\ \hline
RSS         & -5.26          &  -9.08          & \textbf{-19.99}  & 1.05          & \textbf{2.11} & \textbf{9.00} \\ \hline
RSS-WOA     & \textbf{-4.53} &\textbf{ -8.60}  & -63.62          & \textbf{0.62} & 3.21          & 74.66        \\ \hline
DRDPG-WA    & -5.57          & -14.85          & -46.69          & 1.26          & 6.13          & 48.14         \\ \hline
DRDPG-WOA   & -4.54          & -9.76           & -31.24          & 0.91          & 3.24          & 19.68        \\ \hline
DRQN        & -9.31          & -159.68         & -292.67         & 5.68          & 18.01         & 21.85         \\ \hline
Random      & -43.00         & -158.84         & -296.84         & 5.24          & 13.40         & 10.64         \\ \hline
\end{tabular}
\end{table}

Last but not least, as the number of valid actions increases, the performance of DRQN and random policy gradually becomes the same. This is because that DRQN belongs to value-based DRL algorithms, whose core is to learn to approximate the action-values for all actions at each state by utilizing ANNs. However, such an approximation will fail to function if the action space is too large \cite{RL_Introduction}. It is noteworthy that when there are more sensors (e.g., more than 40 sensors) involved, the action space will be too large to be compatible with the centralized scheduling policy and the single-agent DRL algorithm. Then, one possible way to solve the problem is to decompose the joint action space based on multi-agent learning, which is left for future work.

Next, we investigate the computational efficiency and hardware requirements of our proposed RSS algorithm. In Table~\ref{Tab:Time}, we present the average execution time for one forward pass (i.e., action generation), one minibatch training (i.e., replaying and updating), and one simulation (consisting of 500 training episodes and 500 evaluations) for these simulations, which are respectively denoted as ``Forward'', ``One train'', and ``One simulation'' for concise representation.
Furthermore, the required number of parameters, Floating Point Operations (FLOPs) for one forward pass, and storage space are also recorded and presented in Table~\ref{Tab:Time}.
It can be seen that the time spent on one action generation is merely about 0.44 ms in a 7-CSP network\footnote{This latency slightly increases to 0.52 ms if a personal laptop with an Intel(R) Core(TM) i5-8250U@1.60GHz CPU is used. Nevertheless, it also satisfies the real-time control requirement defined by the Open Radio Access Network (O-RAN) Alliance, i.e., below 10 ms \cite{RE_O_RAN_10SM}.}, thereby allowing the system to insert real-time decisions on status updates in practice. Moreover, it is interesting to see that the time consumption for one simulation is roughly the same (about 2.5 hours) when the number of CSPs $K$ varies.
The reason is primarily due to the fact that the dimensions of the observation and generated action linearly increase with respect to the number of CPSs $K$. Consequently, only the number of neurons in the input layer of RSCNs and RSAN, and that in the output layer of RSAN linearly increase in $K$. Therefore, as per Remark~\ref{Remark:Compu_Complexity}, the computational complexity is roughly the same when $K$ rises from 3 to 7.
Finally, it is noteworthy that our proposed RSS algorithm adopts extremely lightweight ANNs, thereby making it less computationally demanding and requiring smaller memory footprints (as shown in Table \ref{Tab:Time}).
% is much less computationally intensive and requires much smaller memory footprints.
As a result, with the rapid development of the hardware (e.g., NVIDIA Jetson AGX Xavier and AGX Orin \cite{AGX_Orin_2022}) and software (e.g., Google Tensorflow Lite \cite{TensorFlow_Lite_2021}) for edge intelligence,  our proposed algorithm is expected to be well compatible with the emerging IoT networks.}

\begin{table}[!t]
\centering
\caption{Evaluation of the computational efficiency and hardware requirements of our proposed RSS algorithm.}
\label{Tab:Time}
\begin{tabular}{|c|c|c|c|}
\hline
\multicolumn{1}{|c|}{$K$} & Forward (ms) & One train (ms) & One simulation     \\ \hline
3                                & 0.43         & 16.48         & 2h 32m 37s           \\ \hline
5                                & 0.44         & 16.58         & 2h 32m 51s          \\ \hline
7                                & 0.44         & 16.76         & 2h 32m 58s           \\ \hline
$K$                       & Parameters   & FLOPs          & Storage space (kB) \\ \hline
3                                & 137,347      & 138,240        & 540                \\ \hline
5                                & 140,677      & 141,568        & 554                \\ \hline
7                                & 144,007      & 144,896        & 568                \\ \hline
\end{tabular}
\end{table}

\begin{table*}[!t]
\centering
\caption{Mean and SD of the average reward achieved by RSS and DRDPG-WA algorithms during the last 50 evaluations.}
\label{Tab:Mean_Std_Conv_EHP}
\begin{tabular}{|c|c|c|c|c|c|c|c|}
\hline
\multirow{2}{*}{\textbf{EH Probability}} &\multirow{2}{*}{\textbf{Algorithm}}
& \multicolumn{3}{c|}{\textbf{Mean}}                & \multicolumn{3}{c|}{\textbf{Standard Deviation}}
\\ \cline{3-8}
       &   & $K=3$   & $K=5$   & $K=7$    & $K=3$  & $K=5$  & $K=7$  \\ \hline
\multirow{2}{*}{0.2}  & RSS      & \textbf{-5.26}  &\textbf{-9.08}   & \textbf{-19.99}  & \textbf{1.05}  & \textbf{2.11}    & \textbf{9.00} \\ \cline{2-8}

	                &DRDPG-WA	 &-5.57	        &-14.85           &-46.69	        &1.26	         &6.13             &48.14  \\ \hline

\multirow{2}{*}{0.4}	&RSS	      &\textbf{-2.32}   &\textbf{-5.68}   &\textbf{-12.13}	   &\textbf{0.26}   &\textbf{2.00}	 &\textbf{22.57}        \\ \cline{2-8}

	                &DRDPG-WA  &-2.76            &-6.84	          &-24.43	        &0.53	         &4.68             &33.98     \\ \hline

\multirow{2}{*}{0.6} 	&RSS	     &\textbf{-1.82}   &\textbf	{-3.17}  &\textbf{-6.40}     &\textbf{0.07}   &\textbf{0.27}     &\textbf{1.55}      \\ \cline{2-8}

	               &DRDPG-WA  &-2.22	       &-4.04	         &-8.00	           &0.57	         &1.32	           &4.44     \\ \hline

\multirow{2}{*}{0.8} &RSS	     &\textbf{-1.71}	 &\textbf{-2.98}   &\textbf{-5.14}	 &\textbf{0.07}	    &\textbf{0.32}    &\textbf{0.58}    \\ \cline{2-8}

	               &DRDPG-WA	&-2.01	      &-3.51            &-6.23             &0.36	         &0.67	           &4.01       \\ \hline

\multirow{2}{*}{1.0}	&RSS	     &\textbf{-1.60}  &\textbf{-2.24}  &\textbf{-3.20}    &\textbf{0.06}	    &\textbf{0.11}	&\textbf{0.34}    \\ \cline{2-8}
	              &DRDPG-WA	&-1.66	      &-2.36	        &-3.73            	&0.17             &0.22	           &0.87   \\ \hline
\end{tabular}
\end{table*}

\subsection{Effectiveness Evaluation}
%
%\begin{figure} [!t]
%\centering
%\leavevmode \epsfxsize=3.0in \epsfbox{change_ehp_3_5_7CSPs.eps}
%\centering \caption{Performance comparison in networks consisting of $K = 3, K = 5$, and $K = 7$ CSPs, where the EH probability varies from 0.2 to 1.0.} \label{Fig:change_ehp_3_5_7CSPs}
%\end{figure}

%\begin{figure} [!t]
%\centering
%\leavevmode \epsfxsize=3.0in \epsfbox{change_K_2_8CSPS.eps}
%\centering \caption{} \label{Fig:change_K_2_8CSPS}
%\end{figure}

%In this subsection, we evaluate the performance of our proposed RSS algorithm in different scenarios. As presented in the previous subsection, incorporating the ADM mechanism can significantly improve the convergence rate of the DRL algorithms, i.e., RSS and DRDPG-WA. As such, to try to ensure that the convergence can be attained in a given training time, we only compare the performance of RSS and DRDPG-WA in different scenarios. Unless otherwise specified, we set the maximum number of training episodes to 500 and calculate the mean/SD of the average reward achieved over the last 50 (451-500) evaluations. All simulation results are obtained by averaging over six independent runs, and the same seed is adopted for both the RSS and DRDPG-WA algorithms in each run.

In this part, we evaluate the performance of our proposed RSS algorithm in different scenarios. The maximum number of training episodes is set to 500 and the mean/SD of the achieved average reward is calculated over the last 50 evaluations.
In Fig. \ref{Fig:change_ehp_3_5_7CSPs}, we compare the performance of RSS and the baseline DRL algorithms (i.e., the DRDPG-WA, RSS-WOA, DRDPG-WOA, and DRQN) in terms of the achieved average reward in three different scenarios, i.e., $K=3$, $K=5$, and $K=7$, where the EH probability varies from 0.2 to 1.0 \cite{EH_Pro_JSAC_2016}. It can be seen that the average reward achieved by DRDPG-WOA and DRQN are low and unstable due to their poor stability and slow convergence rate. Meanwhile, as discussed in the previous subsection, after the convergence is attained, RSS-WOA slightly outperforms RSS in terms of the achieved average reward, while suffering from a much slower convergence rate.

Furthermore, we demonstrate the mean and SD of the average reward achieved by RSS and DRDPG-WA algorithms in Table \ref{Tab:Mean_Std_Conv_EHP}.
By marking the best results in bold, it is readily seen that our proposed RSS algorithm outperforms the baseline DRDPG-WA algorithm in all scenarios, the reason for which is similar to that presented in the previous subsection.
Besides, two other interesting observations were made.
First, for all algorithms, the achieved average reward decreases with the number of CSPs $K$ when the EH probability is given.
This is mainly because, as presented in (\ref{Eq:Succ_Gen_Inte_Info}) and (\ref{Eq:AoI_II_Evolution}), the AoCI is lowered to the minimum value 1 at the end of one time slot, only if for all CSPs the aggregated updates are good enough, the probability of which however decreases with increasing $K$.
Second, for the network with a given number of CSPs, the achieved average reward increases with respect to the EH probability, while the performance of different algorithms gradually becomes the same as the EH probability approaches $1.0$. The main reason is that when the EH rate gradually exceeds the energy consumption rate, the effects of the energy causality constraint and the unobservability of sensors' true battery states on the status update policy design gradually diminish.
As a result, the disparity in the effects of actions on the obtained reward gradually diminishes, thereby mitigating the performance degradation caused by training errors.

\begin{figure} [!t]
\centering
\subfigure[] {\leavevmode \epsfxsize=3.0in  \epsfbox{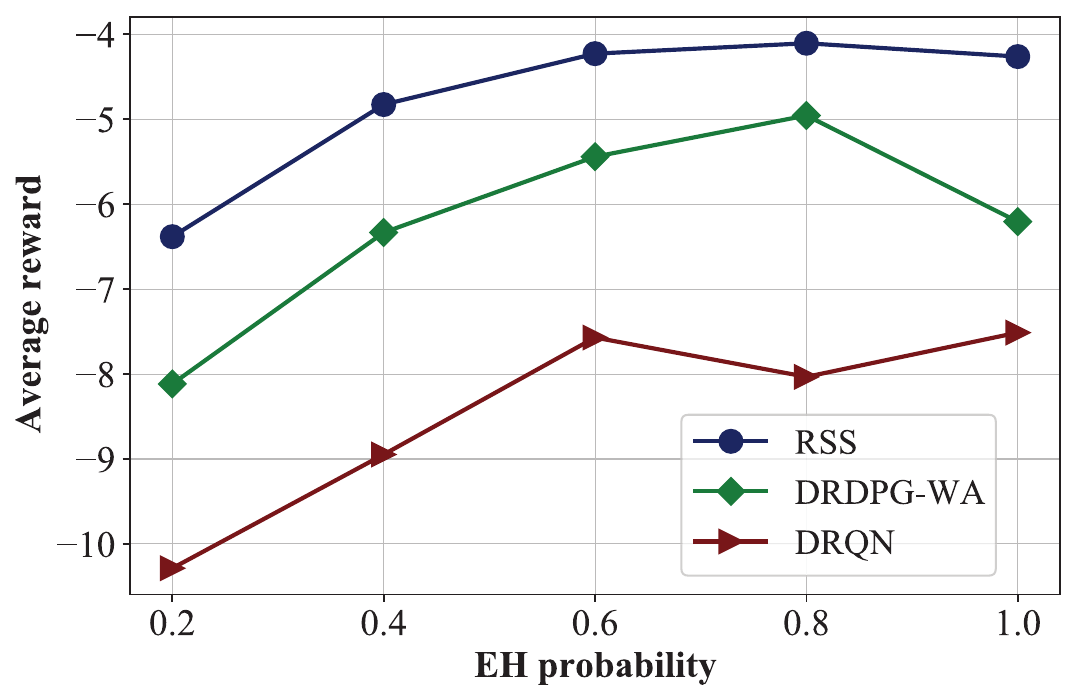}}
\subfigure[] {\leavevmode \epsfxsize=3.0in  \epsfbox{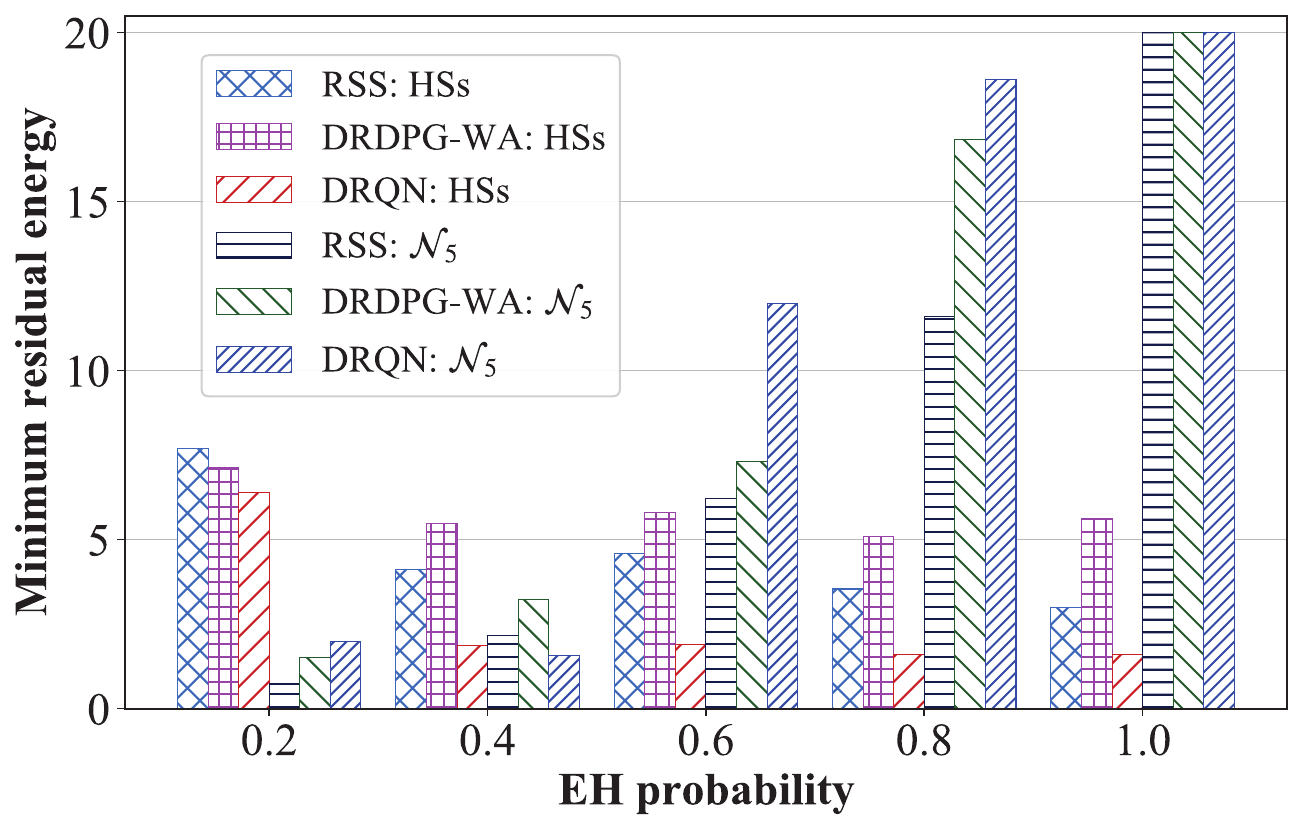}}
\centering \caption{Performance comparison in terms of: (a) the achieved average reward; (b) the average MRE for different sensor sets, with $\left| \mathcal N_5 \right|=3$ and the EH probability of each sensor in $\mathcal N_5$ varying from 0.2 to 1.0.}
\label{Fig:Hetero_EH}
\end{figure}

\begin{figure} [!t]
\centering
\subfigure[] {\leavevmode \epsfxsize=3.05in  \epsfbox{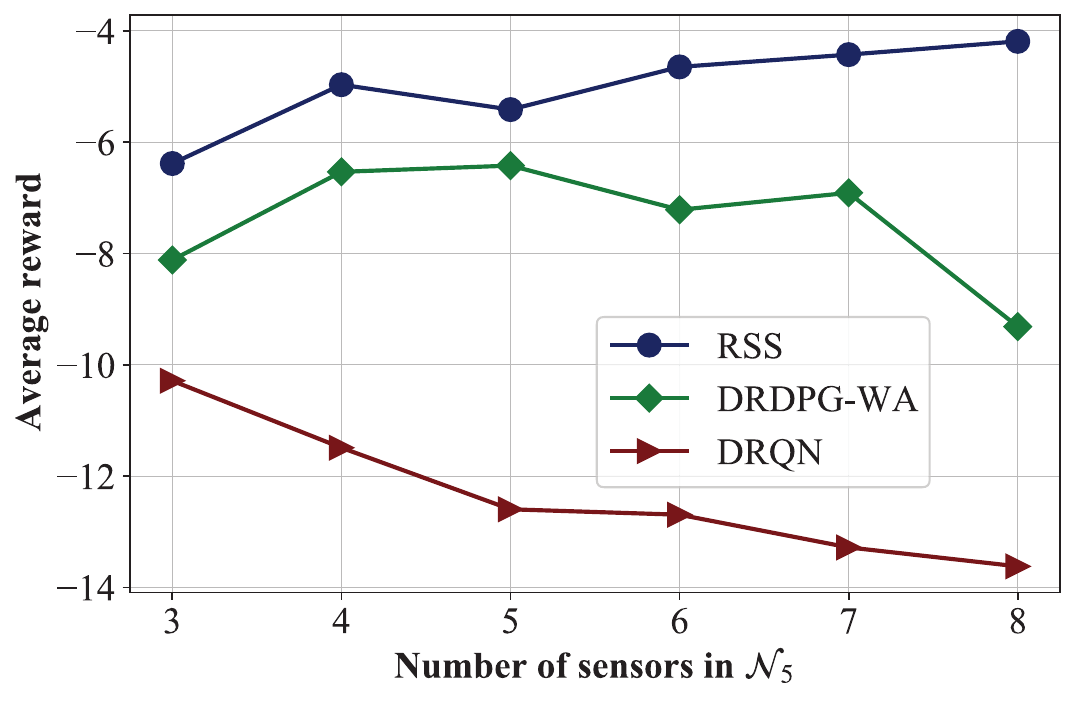}}
\subfigure[] {\leavevmode \epsfxsize=3.0in  \epsfbox{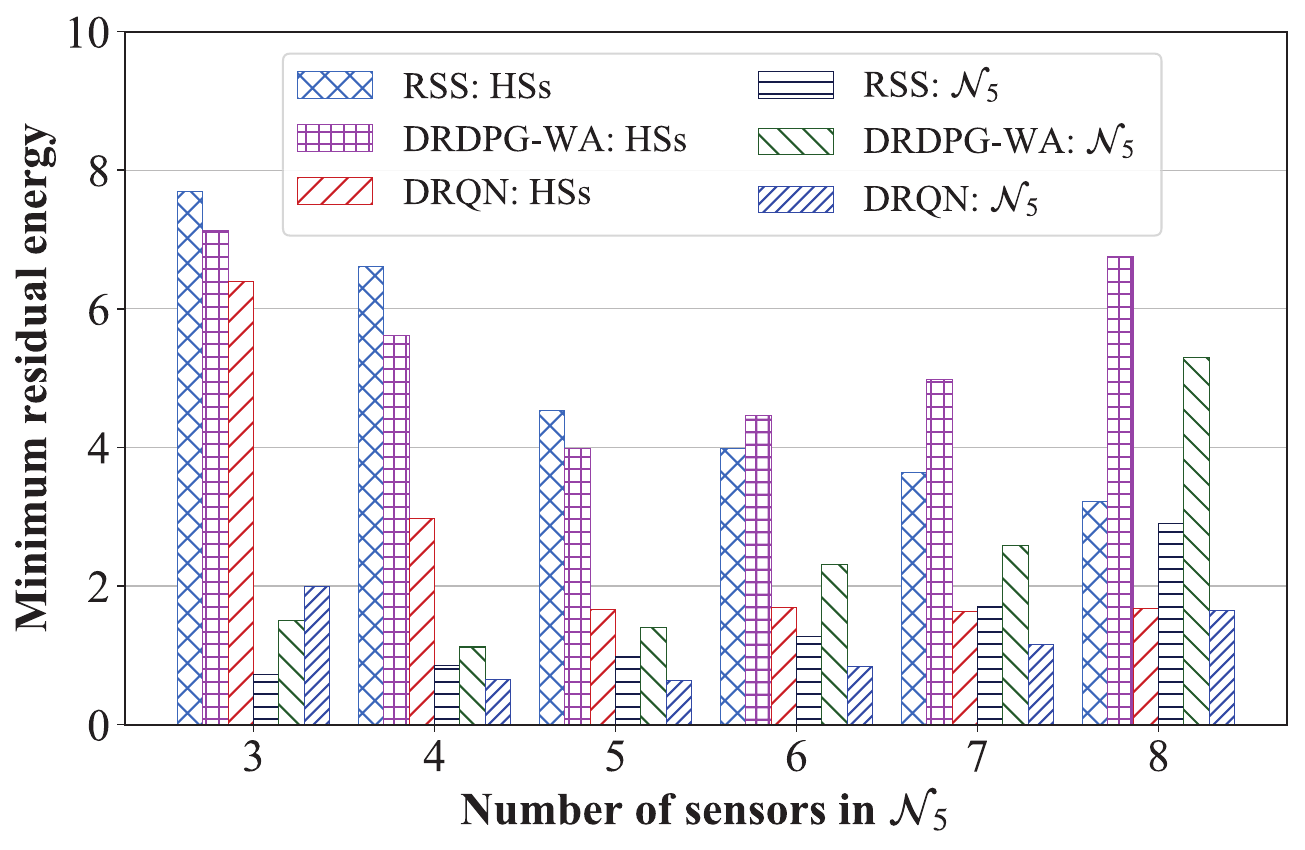}}
\centering \caption{Performance comparison in terms of: (a) the achieved average reward; (b) the average MRE for different sensor sets, with number of sensors in $\mathcal N_5$ varying from 3 to 8.}
\label{Fig:Hetero_Num_Sensor}
\end{figure}

Next, we evaluate the performance of RSS in IoT networks that incorporate heterogeneous sensor sets. Particularly, we consider that there are $K=5$ CSPs, and the sensor sets associated with the first four CSPs (i.e., $\mathcal N_k, \forall k \in \{1, 2, 3, 4\}$) are homogeneous sets (HSs), in which the EH probability is set to 0.4 and other parameters are set according to Section \ref{Sec:Simulation_Setting}. Nevertheless, there is a distinct configuration for $\mathcal N_5$, where the importance of status updates and the transmission failure probability are respectively set to 1.0 and 0.2.
As presented in Figs.~\ref{Fig:Convergence} and \ref{Fig:change_ehp_3_5_7CSPs}, the convergence rate of the DRL-based algorithms can be significantly improved by incorporating the ADM  mechanism, and RSS-WOA only slightly outperforms RSS if the convergence is attained. In this light, for clear exposition, we choose DRDPG-WA and DRQN as baseline algorithms in the remainder of this subsection. The performance comparisons are shown in Figs. \ref{Fig:Hetero_EH} and \ref{Fig:Hetero_Num_Sensor}, where the EH probability and the number of sensors in $\mathcal N_5$ are respectively changed.

% To be specific, we first consider that there are $3$ sensors in the set $\mathcal N_5$ and change the EH probability of each sensor from 0.2 to 1.0.
Fig. \ref{Fig:Hetero_EH} (a) clearly demonstrates the superiority of our proposed RSS algorithm. Particularly, when the EH probability varies from 0.2 to 1.0, compared with DRDPG-WA and DRQN, the RSS's improvement in the achieved average reward ranges from $20.7\%$ to $45.8\%$ and from $61.3\%$ to $95.4\%$, respectively.
It is also interesting to see that the averaged reward achieved by RSS gradually increases with respect to the EH probability, while the trend becomes flat as the EH probability approaches $1.0$.
More specifically, when the EH probability is larger than 0.6, the average rewards achieved by the RSS algorithm are roughly the same, with the fluctuation caused by the randomness in the simulation.
And the sudden drop in the curve of DRDPG-WA occurs because the convergence is not completely achieved.
Such an increasing trend is mainly attributed to the fact that, in each HS there are $4$ sensors with the EH probability set as 0.4, while only 3 sensors are deployed to monitor CSP 5.
As such, when the EH probability is small (e.g., 0.2 and 0.4), the achieved average reward is mainly constrained by the available energy of sensors in $\mathcal N_5$.
Nevertheless, as the EH probability increases, the status update capability of sensors in $\mathcal N_5$ is enhanced, and the available energy of sensors in HSs gradually becomes the bottleneck for improving the information freshness.
To verify this, we record the minimum residual energy (MRE) of sensors in each set at the beginning of each time slot, and present the average MRE for HSs and $\mathcal N_5$ in Fig. \ref{Fig:Hetero_EH} (b). As illustrated in Fig. \ref{Fig:Hetero_EH} (b), the average MRE in $\mathcal N_5$ monotonically increases with respect to the EH probability, and reaches the battery capacity (i.e., 20 units) when the EH probability equals one.

In Figs. \ref{Fig:Hetero_Num_Sensor} (a) and (b), the achieved average reward and the average MRE for HSs and $\mathcal N_5$ are respectively presented, in which the number of sensors in $\mathcal N_5$ varies from 3 to 8. In this context, similar observations can be made as in Fig. \ref{Fig:Hetero_EH}, except that, on the one hand, the superiority of RSS to DRQN is more pronounced when the number of sensors in $\mathcal N_5$ increases (which results in an expanded action space).
Here, the small fluctuation at $\left| \mathcal N_5 \right|=5$ originates from the randomness in the simulation.
On the other hand, by implementing our proposed RSS algorithm, the average MRE for HSs and $\mathcal N_5$ tends to be the same with the increasing size of $\mathcal N_5$.
This is mainly due to the fact the EH probability of each sensor in HSs is greater than that in set $\mathcal N_5$ and, to improve the AoCI at the DFC, the energy harvested by sensors in different sets needs to be synergistically managed. By learning the environmental dynamics and making well-informed decisions, RSS can adaptively schedule the sensors in different sets to improve the energy utilization and reduce the AoCI.
In contrast, the implementation of DRQN leads to both the lowest average reward and nearly the smallest average MRE.
This phenomenon emphasizes the significance of appropriately managing the energy harvested by sensors. That is if the EH sensors cannot be adaptively scheduled the harvested energy cannot be efficiently used.

\section{Conclusions}
This work considered an IoT network with groups of EH sensors deployed to simultaneously monitor multiple CSPs.
We aimed to minimize the AoCI at the DFC by adaptively scheduling the sensors to send timely status updates, where the sensors' true battery states are unobservable to the DFC when making decisions.
We formulated the dynamic status update procedure as a POMDP and developed an RNN-enhanced DRL algorithm to solve it.
By leveraging the advantages of the SAC and LSTM techniques, our proposed RSS algorithm has successfully tackled the challenges arising from the causality of energy usage, unknown environmental dynamics, unobservability of sensors' battery states, and large-scale discrete action space.
Extensive simulations have been conducted, showing that our proposed RSS algorithm surpasses the DRQN- and DRDPG-based status update algorithms in terms of convergence, scalability, and stability regarding the achieved average reward, i.e., the negative of the average AoCI.
More importantly, by incorporating our devised ADM mechanism, the RSS algorithm can achieve convergence even in the large-scale scenario with more than $800,000$ valid actions.

\begin{figure*} [!t]
\begin{align} \label{Eq:Action_Value_Fun_Reform}
\nonumber & Q_{\pi}(\mathcal Z(t),  \mathbf A(t)) =  \sum\nolimits_{U\in\mathcal U, \mathcal S' \in \mathbb S,  \mathcal O' \in \mathbb O} {\mathsf {Pr}_{(\mathcal S(t), \mathbf A(t))}^{(\mathcal S', U, \mathcal O')}} \left(U + \mathbb E \big[\sum\nolimits_{l = 1}^\infty  \gamma^{l}\left(U\left( {{\cal S}(t + l),{ \mathbf A}(t + l)}\right) + \alpha \mathcal H \left(  \pi \left(\cdot  \left| \mathcal Z(t+l) \right.\right) \right)\right)\big]\right) \\
&=\sum\nolimits_{U\in\mathcal U, \mathcal S' \in \mathbb S,  \mathcal O' \in \mathbb O} {\mathsf {Pr}_{(\mathcal S(t), \mathbf A(t))}^{(\mathcal S', U, \mathcal O')}} \left(U + \gamma \mathbb E \big[\sum\nolimits_{l = 0}^\infty  \gamma^{l}\left(U\left( {{\cal S}(t' + l),{ \mathbf A}(t' + l)}\right) + \alpha \mathcal H \left(  \pi \left(\cdot  \left| \mathcal Z(t'+l) \right.\right) \right)\right)\big]\right)
\end{align}
\hrulefill
\end{figure*}

\setcounter{equation}{35}

\begin{figure*} [!t]
\begin{align} \label{Eq:State_Value_Fun_Bell}
V_{\pi}(\mathcal Z(t)) &= \mathbb E_{\mathbf A\sim\pi} \big[\sum\nolimits_{U\in\mathcal U, \mathcal S' \in \mathbb S,  \mathcal O' \in \mathbb O} {\mathsf {Pr}_{(\mathcal S(t), \mathbf A(t))}^{(\mathcal S', U, \mathcal O')}} \left(U + \alpha \mathcal H \left(  \pi \left(\cdot  \left| \mathcal Z(t) \right.\right) \right)+\gamma V_{\pi}(\mathcal Z')  \right)\big]  \\ \nonumber
& \mathop  = \limits^{(a)} \mathbb E_{\mathbf A\sim\pi} \big[\left(\sum\nolimits_{U\in\mathcal U, \mathcal S' \in \mathbb S,  \mathcal O' \in \mathbb O} {\mathsf {Pr}_{(\mathcal S(t), \mathbf A(t))}^{(\mathcal S', U, \mathcal O')}} \left(U  +\gamma V_{\pi}(\mathcal Z')  \right)\right)- \alpha\log \pi (\mathbf A  \left| \mathcal Z(t) \right.)\big]
\end{align}
\hrulefill
\end{figure*}

An interesting extension of this work is to design distributed dynamic status update policies to minimize the AoCI at the DFC, enabling individual sensors to learn to make well-informed decisions.
In this case, each sensor can observe its own battery state, while how to accurately evaluate the long-term effects of its own actions on the achieved average AoCI and design efficient multi-agent DRL-based status update algorithms becomes the main concern.
In addition, this paper assumed the importance of data updated by each individual sensor was fixed.
By incorporating the data processing (e.g., data compression) scheme and properly modeling the associated energy consumption, the model can be extended to allow the DFC to dynamically control the importance of updates generated by the sensors. This capability would be crucial for facilitating semantic-aware and goal-oriented communications \cite{Availa_MC_2022, RE_O_RAN_10SM}.
In this scenario, the data processing and sensor scheduling need to be jointly optimized, and a POMDP with both continuous (w.r.t. data compression) and discrete (w.r.t. sensor scheduling) actions shall be formulated and solved.

%\ifCLASSOPTIONcompsoc
%  % The Computer Society usually uses the plural form
%\else
%  % regular IEEE prefers the singular form
%\fi
%
%This paper was supported by the Chinese Universities Scientific Fund (2452017560), High-level Talents Fund of Shaanxi Province (F2020221001), and the Technological Innovation Fund of Shaanxi Academy of Forestry (SXLK2021-0215).
%
%
%\ifCLASSOPTIONcaptionsoff
%  \newpage
%\fi

%\clearpage

\appendices
\section{Proof of Theorem \ref{The:Consis_Ccond}} \label{Pro:Lemm_Consis_Ccond}
\begin{IEEEproof}
According to the definition of the SA function $ Q_{\pi}(\mathcal Z(t),  \mathbf A(t))$ presented in (\ref{Eq:Act_Value_Fun}), we can transform its expression as shown in (\ref{Eq:Action_Value_Fun_Reform}) by equivalently replacing $t+1$ with $t'$, $\mathcal S'$ with $\mathcal S(t')$, and $\{\mathcal Z(t)\}\bigcup \mathcal O'$ with $\mathcal Z(t')$. Wherein, for notation simplicity, we denote the probability of the transition to state $\mathcal S'$ with reward $U$ and observation $\mathcal O'$ from the state-action pair $(\mathcal S, \mathbf A)$ by ${\mathsf {Pr}_{(\mathcal S, \mathbf A)}^{(\mathcal S', U, \mathcal O')}}$, i.e., $\forall S, S' \in \mathbb S, \forall \mathbf A \in \mathbb A, \forall U \in \mathcal U, \forall \mathcal O' \in \mathbb O$,
\setcounter{equation}{34}
\begin{align} \label{Eq:Trans_Prob}
\mathsf {Pr}_{(\mathcal S, \mathbf A)}^{(\mathcal S', U, \mathcal O')} = {\mathsf {Pr}\left( {{U, \mathcal S', \mathcal O'}\left| {{\mathcal S},{\mathbf A}} \right.} \right)}.
\end{align}
Then, by substituting (\ref{Eq:State_Value_Fun}) into (\ref{Eq:Action_Value_Fun_Reform}) and replacing $\mathcal Z(t')$ with $\mathcal Z'$, we can derive the recursive equation shown in (\ref{Eq:Action_Value_Fun_Ite}).

\setcounter{equation}{36}

On the other hand, following (\ref{Eq:State_Value_Fun}) we can express the Bellman equation \cite{RL_Introduction} for the SS function as shown in (\ref{Eq:State_Value_Fun_Bell}) with $\mathcal Z' = \{\mathcal Z(t)\}\bigcup \mathcal O'$, where (a) holds since $\mathcal H \left(  \pi (\cdot  \left| \mathcal Z(t) \right.) \right)$ is already an expectation over policy $\pi$ \cite{Entropy_Shannon_2001}, i.e.,
\begin{align}
\nonumber \mathcal H \left(  \pi (\cdot  \left| \mathcal Z(t) \right.) \right)& = \sum\nolimits_{\mathbf A \in \mathbb A} -\pi (\mathbf A  \left| \mathcal Z(t) \right.)\log\pi (\mathbf A  \left| \mathcal Z(t) \right.) \\
&=\mathbb E_{\mathbf A\sim\pi} \big[ -\log\pi (\mathbf A  \left| \mathcal Z(t) \right.) \big].
\end{align}
Finally, by substituting the recursive equation (\ref{Eq:Action_Value_Fun_Ite}) into (\ref{Eq:State_Value_Fun_Bell}) we can draw the conclusion as shown in (\ref{Eq:State_Value_Fun_Ite}).
\end{IEEEproof}

\bibliographystyle{IEEEtran}
%% argument is your BibTeX string definitions and bibliography database(s)
\bibliography{IEEEabrv,RL_UP_Ref}

\begin{IEEEbiography}[{\includegraphics[width=1in,height=1.25in,clip,keepaspectratio]{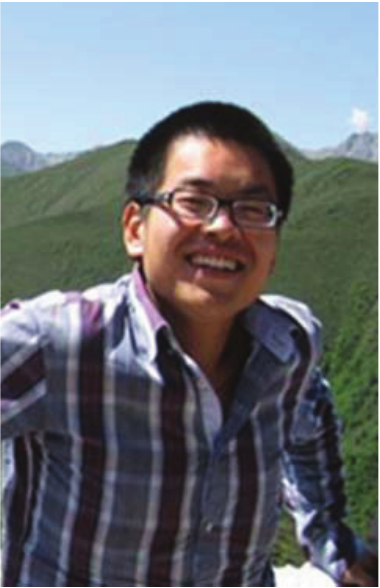}}]
{Chao Xu} (M'15)
earned the Ph.D. degree in information and communication engineering from Xidian University, Xi'an, China, in 2015. He is currently a Full Professor with the School of Information Engineering, Northwest A\&F University, Yangling, China, and employed as a researcher with the Key Laboratory of Agricultural Internet of Things, Ministry of Agriculture and Rural Affairs, Yangling. His current research interests include AoI analysis and optimization, and machine learning for wireless communications and Internet of Things (IoT) networks.
\end{IEEEbiography}

%\begin{IEEEbiography}[{\includegraphics[width=1in,height=1.25in,clip,keepaspectratio]{Chao_Xu.eps}}]
%{Chao Xu} (M'15)
%received the B.S. degree in electronic information engineering and the Ph.D. degree in information and communication engineering from Xidian University, Xi'an, China, in 2009 and 2015, respectively. He was a Postdoctoral Researcher from 2015 to 2017 with the School of Telecommunications Engineering, Xidian University. He is currently an Associate Professor with the School of Information Engineering, Northwest A\&F University, Yangling, China, and employed as a researcher with the Key Laboratory of Agricultural Internet of Things, Ministry of Agriculture and Rural Affairs, Yangling. His current research interests include AoI analysis and optimization, and machine learning for wireless communications and Internet of Things (IoT) networks.
%\end{IEEEbiography}

\begin{IEEEbiography}[{\includegraphics[width=1in,height=1.25in,clip,keepaspectratio]{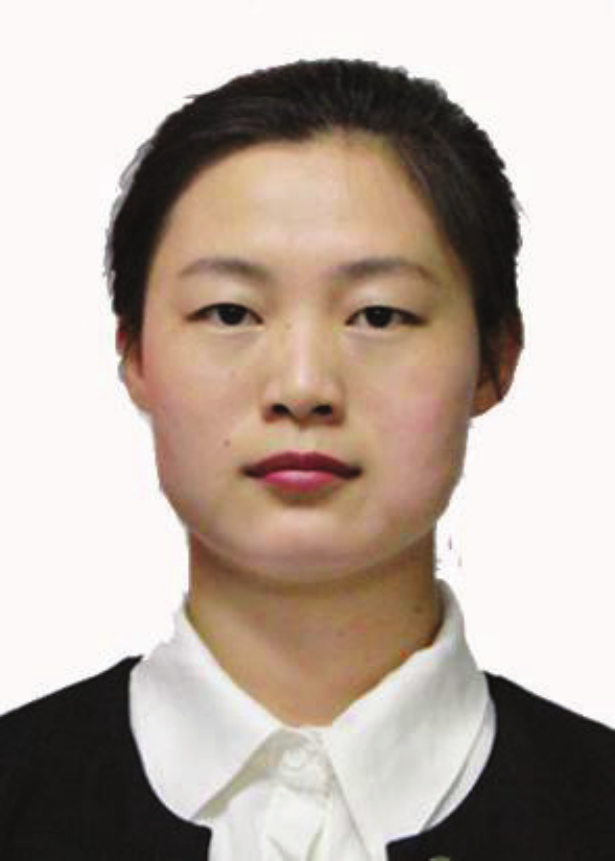}}]
{Xinyan Zhang} received her B.S. degree in information and computing sciences from Northwest A\&F University, Yangling, China, in 2018. Currently, she is pursuing the master degree in software engineering at Northwest A\&F University, Yangling, China. Her research interest is the application of deep reinforcement learning for dynamic control in Internet of Things (IoT) networks.
\end{IEEEbiography}

\begin{IEEEbiography}[{\includegraphics[width=1in,height=1.25in,keepaspectratio]{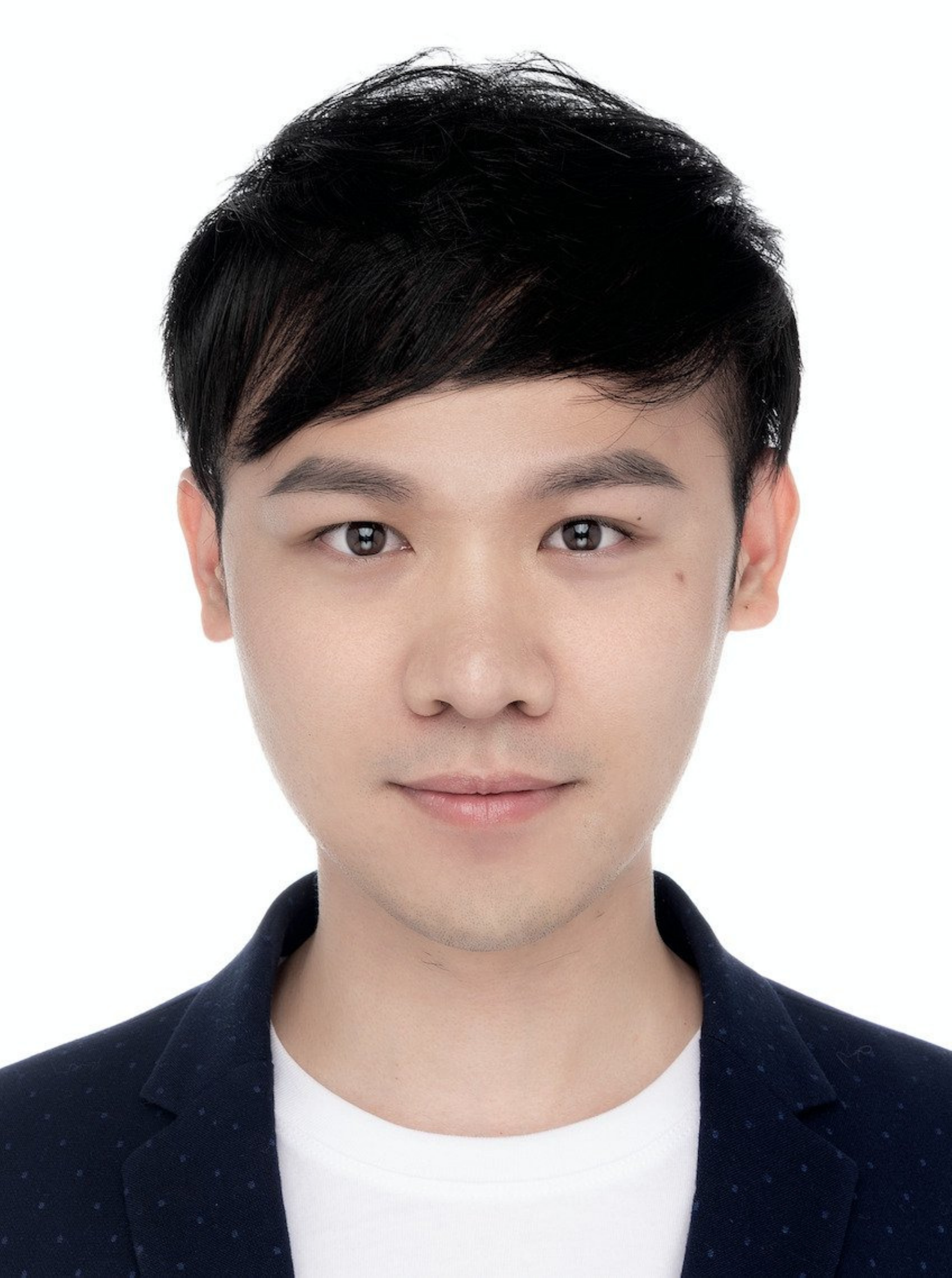}}]
{Howard H. Yang}(S'13-M'17) earned the Ph.D. degree in Electrical Engineering from Singapore University of Technology and Design (SUTD), Singapore, in 2017. Currently, he is an assistant professor with the Zhejiang University/University of Illinois at Urbana-Champaign Institute (ZJU-UIUC Institute), Zhejiang University, Haining, China.
Dr. Yang currently serves as an editor for the {\scshape IEEE Transactions on Wireless Communications}. His research interests cover various aspects of wireless communications, networking, and signal processing, currently focusing on the modeling of modern wireless networks, high dimensional statistics, graph signal processing, and machine learning.
\end{IEEEbiography}

\begin{IEEEbiography}[{\includegraphics[width=1in,height=1.25in,clip,keepaspectratio]{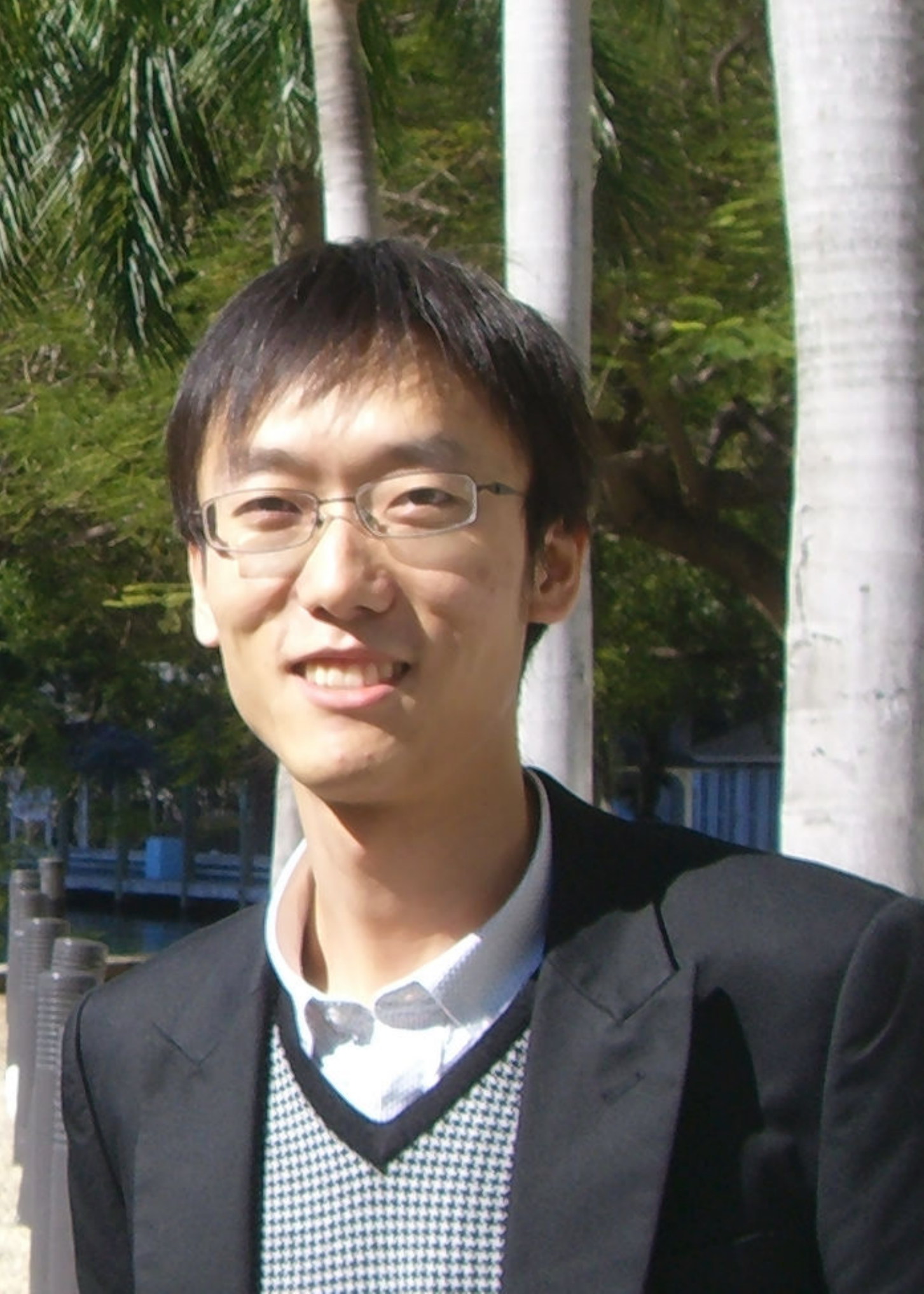}}]
{Xijun Wang} (M'12)
earned the Ph.D. degree in Electronic Engineering from Tsinghua University in January 2012, Beijing, China. He was an Assistant Professor from 2012 to 2015 and an Associate Professor from 2015 to 2018 with the School of Telecommunications Engineering, Xidian University. He is currently an Associate Professor with Sun Yat-sen University. His current research interests include UAV communications, edge computing, and age of information.
\end{IEEEbiography}

\begin{IEEEbiography}[{\includegraphics[width=1in,height=1.25in,keepaspectratio]{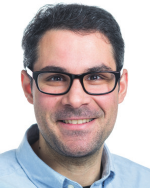}}]
{Nikolaos Pappas} (S'07-M'13-SM'21) earned the Ph.D. degrees in computer science from the University of Crete, Greece, in 2012. He received a B.Sc. degree in mathematics from the University of Crete in 2012. He is currently an Associate Professor at the Department of Computer and Information Science at Link\"{o}ping University, Link\"{o}ping, Sweden. He serves as an Editor for five IEEE journals. His main research interests include the field of wireless communication networks with an emphasis on semantics-aware communications, energy harvesting networks, network-level cooperation, age of information, and stochastic geometry.
\end{IEEEbiography}

\begin{IEEEbiography}[{\includegraphics[width=1in,height=1.25in,keepaspectratio]{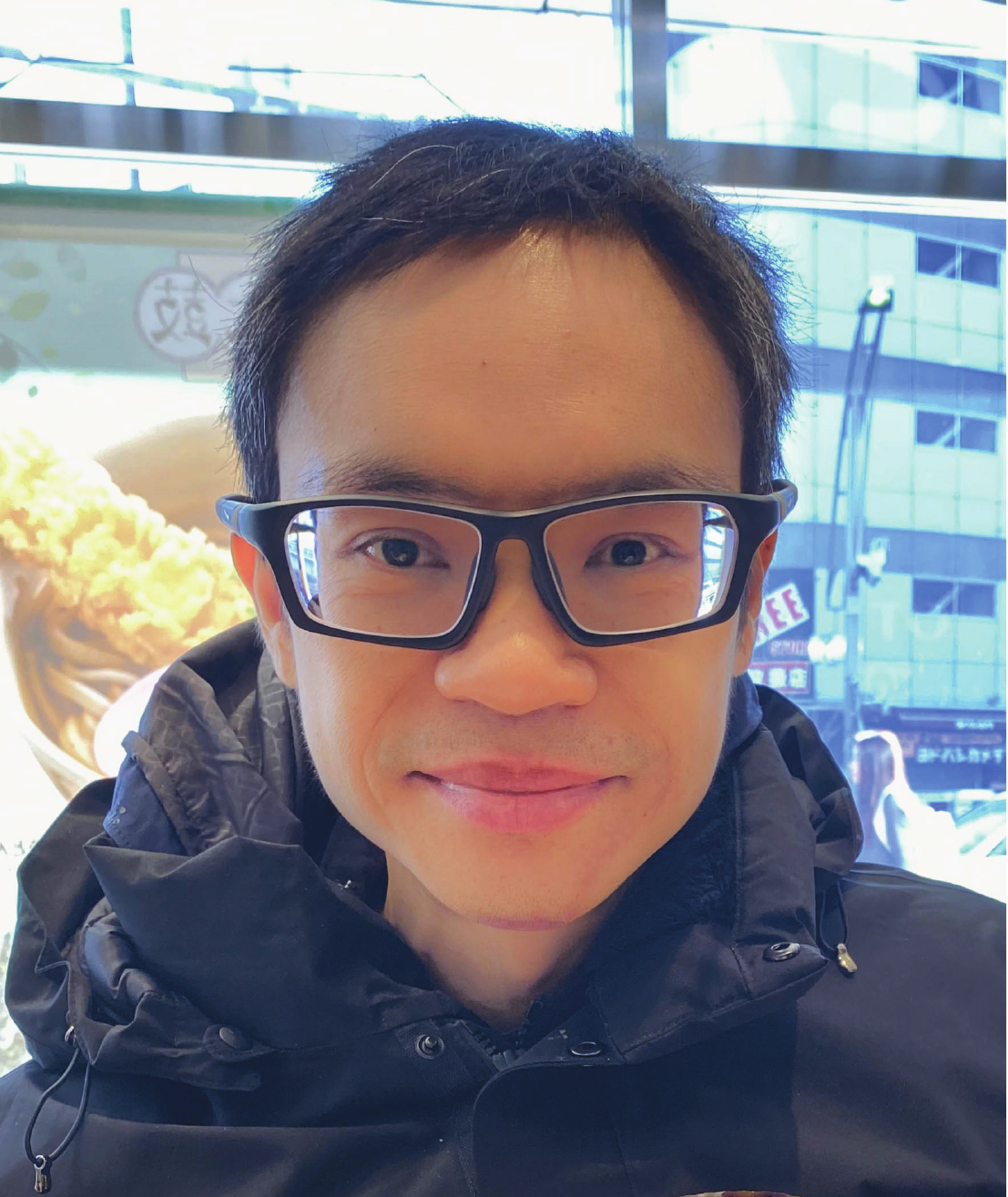}}]
{Dusit Niyato}(M'09-SM'15-F'17) is currently a professor in the School of Computer Science and Engineering, at Nanyang Technological University, Singapore. He received B.Eng. from King Mongkuts Institute of Technology Ladkrabang (KMITL), Thailand in 1999 and Ph.D. in Electrical and Computer Engineering from the University of Manitoba, Canada in 2008. His research interests are in the areas of the Internet of Things (IoT), machine learning, and incentive mechanism design.
\end{IEEEbiography}

\begin{IEEEbiography}[{\includegraphics[width=1in,height=1.25in,keepaspectratio]{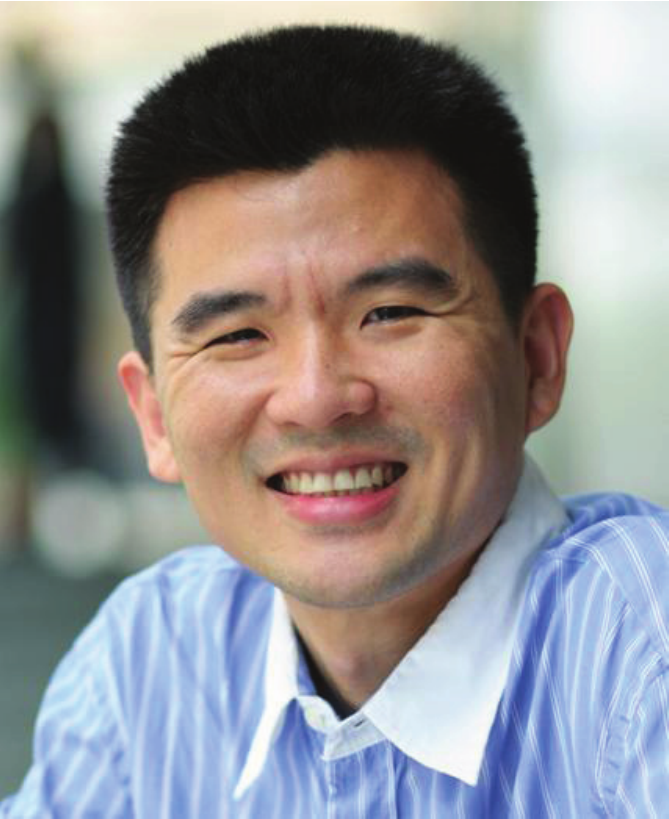}}]
 {Tony Q. S. Quek}(S'98-M'08-SM'12-F'18) earned the Ph.D. degree in electrical engineering and computer science from the Massachusetts Institute of Technology in 2008. Currently, he is the Cheng Tsang Man Chair Professor with Singapore University of Technology and Design (SUTD) and ST Engineering Distinguished Professor. He also serves as the Director of the Future Communications R\&D Programme, the Head of ISTD Pillar, and the Deputy Director of the SUTD-ZJU IDEA.
 Dr. Quek is currently serving as an Area Editor for the {\scshape IEEE Transactions on Wireless Communications}. He is a Fellow of IEEE and a Fellow of the Academy of Engineering Singapore. His current research topics include wireless communications and networking, network intelligence, non-terrestrial networks, open radio access network, and 6G.
\end{IEEEbiography}

\end{document}